# A Novel Stretch Energy Minimization Algorithm for Equiareal Parameterizations

Mei-Heng Yueh · Wen-Wei Lin ·
Chin-Tien Wu · Shing-Tung Yau

**Abstract** Surface parameterizations have been widely applied to computer graphics and digital geometry processing. In this paper, we propose a novel stretch energy minimization (SEM) algorithm for the computation of equiareal parameterizations of simply connected open surfaces with a very small area distortion and a highly improved computational efficiency. In addition, the existence of nontrivial limit points of the SEM algorithm is guaranteed under some mild assumptions of the mesh quality. Numerical experiments indicate that the efficiency, accuracy, and robustness of the proposed SEM algorithm outperform other state-of-the-art algorithms. Applications of the SEM on surface remeshing and surface registration for simply connected open surfaces are demonstrated thereafter. Thanks to the SEM algorithm, the computations for these applications can be carried out efficiently and robustly.

## 1 Introduction

A parameterization of a surface $\mathcal{M} \subset \mathbb{R}^3$ is a bijective mapping $f$ that maps $\mathcal{M}$ to a planar region $\mathcal{D} \subset \mathbb{C}(=\mathbb{R}^2)$. It has been widely applied to various tasks of

Mei-Heng Yueh ⊠ · Wen-Wei Lin · Chin-Tien Wu
Department of Applied Mathematics, National Chiao Tung University, Hsinchu 300, Taiwan
E-mail: mhyueh@math.nctu.edu.tw

Wen-Wei Lin
E-mail: wwlin@math.nctu.edu.tw

Chin-Tien Wu
E-mail: ctw@math.nctu.edu.tw

Shing-Tung Yau
Department of Mathematics, Harvard University, Cambridge, MA 02138, USA
E-mail: yau@math.harvard.edu

digital geometry processing, such as surface registration, resampling, remeshing and texture mapping. Some classical methods and applications of surface parameterizations can be found in the survey papers [11,17,13].

A desired parameterization usually minimizes the distortion of either angles or areas. In particular, an area-preserving mapping is also called an equiareal parameterization. Some popular approaches for the computation of equiareal parameterizations are:

1. stretch-minimizing method [16,20],
2. Lie advection method [24],
3. optimal mass transportation method [9,12,23,18].

In the applications of surface resampling and remeshing, a suitable parameterization can be applied so that the sampling on the surface in 3D space can be determined by the sampling on a planar domain of simple shapes via the one-to-one correspondence of the parameterization mapping. The distortion of the sampling density between the samplings on the surface and on the planar domain, respectively, is caused by the stretch of the parameterization. In order to control the density of the sampling on the surface via a planar domain, a parameterization with a very small stretch is desired. One of the best options is the equiareal parameterization, which refers to as a bijective area-preserving mapping $f$ that maps a surface $\mathcal{M} \subset \mathbb{R}^3$ onto a planar region $\mathcal{D} \subset \mathbb{C}(= \mathbb{R}^2)$.

In computations, the surface we considered is a triangular mesh, i.e., a piecewise linear surface composed of triangles. A triangular mesh $\mathcal{M}$ in $\mathbb{R}^3$ is composed of triangular faces

$$\mathcal{F}(\mathcal{M}) = \{[\mathbf{v}_i, \mathbf{v}_j, \mathbf{v}_k]\},$$

where, $\mathbf{v}_i \equiv \left(\mathbf{v}_i^1, \mathbf{v}_i^2, \mathbf{v}_i^3\right)^\top \in \mathbb{R}^3$ is a vertex of $\mathcal{M}$, the bracket $[\mathbf{v}_1, \ldots, \mathbf{v}_m]$ denotes the *convex hull* of the affinely independent points $\{\mathbf{v}_1, \ldots, \mathbf{v}_m\} \subset \mathbb{R}^3$ defined by

$$[\mathbf{v}_1, \ldots, \mathbf{v}_m] = \left\{ \sum_{\ell=1}^m \alpha_\ell \mathbf{v}_\ell \,\middle|\, \sum_{\ell=1}^m \alpha_\ell = 1, \alpha_\ell \geq 0, \ell = 1, \ldots, m \right\} \subset \mathbb{R}^3.$$

For convenience, we denote the set of vertices of $\mathcal{M}$ by $\mathcal{V}(\mathcal{M})$ and the set of edges of $\mathcal{M}$ by

$$\mathcal{E}(\mathcal{M}) = \{[\mathbf{v}_i, \mathbf{v}_j] \mid [\mathbf{v}_i, \mathbf{v}_j, \mathbf{v}_k] \in \mathcal{F}(\mathcal{M}) \text{ for some } \mathbf{v}_k \in \mathcal{V}(\mathcal{M})\}.$$

### 1.1 Contributions

In this paper, we propose a novel stretch energy minimization (SEM) algorithm for the computation of disk-shaped equiareal parameterizations of simply connected open surfaces with small area distortions and highly improved computational efficiencies. The contributions can be divided into three folds. First, we improve the area distortions of the parameterization mapping on a given surface $\mathcal{M}$ by introducing a novel iterative scheme of the SEM algorithm. The desired equiareal parameterization of $\mathcal{M}$ is achieved as the iteration converges. Second, the accuracy, efficiency and robustness of the proposed SEM algorithm are highly improved compared with other state-of-the-art algorithms. Third, we prove the existence of nontrivial (nonconstant) limit points of the SEM algorithm under some mild assumptions of the mesh quality.

1.2 Notations and Overview

The following notations are frequently used in this paper. Other notations will be clearly defined whenever they appear.

- Bold letters, e.g. $\mathbf{u}$, $\mathbf{v}$, $\mathbf{w}$, denote (complex) vectors.
- Capital letters, e.g. $A$, $B$, $C$, denote matrices.
- Typewriter letters, e.g. $\mathtt{I}$, $\mathtt{J}$, $\mathtt{K}$, denote ordered sets of indices.
- $\mathbf{v}_i$ denotes the $i$-th entry of the vector $\mathbf{v}$.
- $\mathbf{v}_\mathtt{I}$ denotes the subvector of $\mathbf{v}$ composed of $\mathbf{v}_i$, for $i \in \mathtt{I}$.
- $|\mathbf{v}|$ denotes the vector with the $i$-th entry being $|\mathbf{v}_i|$.
- $\mathrm{diag}(\mathbf{v})$ denotes the diagonal matrix with the $(i,i)$-th entry being $\mathbf{v}_i$.
- $A_{i,j}$ denotes the $(i,j)$-th entry of the matrix $A$.
- $A_{\mathtt{I},\mathtt{J}}$ denotes the submatrix of $A$ composed of $A_{i,j}$, for $i \in \mathtt{I}$ and $j \in \mathtt{J}$.
- $[\mathbf{u}, \mathbf{v}, \mathbf{w}]$ denotes the triangular face with the vertices $\{\mathbf{u}, \mathbf{v}, \mathbf{w}\} \subset \mathbb{R}^3$ or $\mathbb{C}$.
- $|[\mathbf{u}, \mathbf{v}, \mathbf{w}]|$ denotes the area of the triangular face $[\mathbf{u}, \mathbf{v}, \mathbf{w}]$.
- $\mathrm{i}$ denotes the imaginary unit $\sqrt{-1}$.
- $n_A$ denotes the number of rows of a squared matrix $A$.
- $n_\mathtt{I}$ denotes the number of elements of an ordered index set $\mathtt{I}$.
- $I_n$ denotes the identity matrix of size $n \times n$.
- $\mathbf{1}_n$ denotes the vector of length $n$ with all entries being one.
- $\mathbf{0}$ denotes the zero vectors and matrices of appropriate sizes.

This paper is organized as follows. First, the discrete equiareal mapping and a brief review of the most related work on the fast stretch minimization are introduced in Sect. 2 and 3, respectively. Then the SEM algorithm for computing equiareal parameterizations of simply connected open surfaces is proposed in Sect. 4. The proof of the existence of nontrivial limit points for the SEM algorithm is given in Sect. 5. Numerical results of the SEM algorithm compared with two state-of-the-art methods [20, 18] are presented in Sect. 6. Finally, some applications of the SEM algorithm on surface remeshing and surface registration are demonstrated in Sect. 7. A concluding remark is given in Sect. 8.

## 2 Discrete Equiareal Mappings

Given a triangular mesh $\mathcal{M}$ of $n$ vertices $\{\mathbf{v}_\ell\}_{\ell=1}^n$ with $\mathbf{v}_\ell \in \mathbb{R}^3$. A bijective piecewise affine mapping $f$ from $\mathcal{M}$ to a unit disk $\mathbb{D}$: $f : \mathcal{M} \to \mathbb{D} \subset \mathbb{C}$ can be expressed by a complex-valued vector

$$\mathbf{f} = (\mathbf{f}_1, \ldots \mathbf{f}_n)^\top \in \mathbb{C}^n,$$

where $\mathbf{f}_\ell = f(\mathbf{v}_\ell)$, for $\ell = 1, \ldots, n$. For convenience, we write $\mathbf{f}_\ell = \mathbf{f}_\ell^1 + \mathrm{i}\mathbf{f}_\ell^2$, where $\mathbf{f}_\ell^1$, $\mathbf{f}_\ell^2 \in \mathbb{R}$, for $\ell = 1, \ldots, n$. For $z \in [\mathbf{f}_i, \mathbf{f}_j, \mathbf{f}_k] \subset \mathbb{D}$, the inverse mapping $f^{-1} : \mathbb{D} \to \mathcal{M}$ is defined by

$$f^{-1}|_{[\mathbf{f}_i, \mathbf{f}_j, \mathbf{f}_k]}(z) = \frac{|[z, \mathbf{f}_j, \mathbf{f}_k]|\mathbf{v}_i + |[\mathbf{f}_i, z, \mathbf{f}_k]|\mathbf{v}_j + |[\mathbf{f}_i, \mathbf{f}_j, z]|\mathbf{v}_k}{|[\mathbf{f}_i, \mathbf{f}_j, \mathbf{f}_k]|},$$

where $|[\mathbf{f}_i, \mathbf{f}_j, \mathbf{f}_k]|$ denotes the area of the triangle element $[\mathbf{f}_i, \mathbf{f}_j, \mathbf{f}_k]$ defined as

$$|[\mathbf{f}_i, \mathbf{f}_j, \mathbf{f}_k]| = \frac{1}{2}\left|(\mathbf{f}_i^1 - \mathbf{f}_k^1)(\mathbf{f}_j^2 - \mathbf{f}_k^2) - (\mathbf{f}_i^2 - \mathbf{f}_k^2)(\mathbf{f}_j^1 - \mathbf{f}_k^1)\right|.$$

Let $z = z^1 + \mathrm{i} z^2$. The partial derivatives of $f^{-1}$ with respect to $z^1$ and $z^2$ are expressed by

$$\frac{\partial f^{-1}|_{[\mathbf{f}_i, \mathbf{f}_j, \mathbf{f}_k]}}{\partial z^1}(z) = \frac{1}{2|[\mathbf{f}_i, \mathbf{f}_j, \mathbf{f}_k]|} \left[ (\mathbf{f}_j^2 - \mathbf{f}_k^2)\mathbf{v}_i + (\mathbf{f}_k^2 - \mathbf{f}_i^2)\mathbf{v}_j + (\mathbf{f}_i^2 - \mathbf{f}_j^2)\mathbf{v}_k \right]$$

and

$$\frac{\partial f^{-1}|_{[\mathbf{f}_i, \mathbf{f}_j, \mathbf{f}_k]}}{\partial z^2}(z) = \frac{1}{2|[\mathbf{f}_i, \mathbf{f}_j, \mathbf{f}_k]|} \left[ (\mathbf{f}_k^1 - \mathbf{f}_j^1)\mathbf{v}_i + (\mathbf{f}_i^1 - \mathbf{f}_k^1)\mathbf{v}_j + (\mathbf{f}_j^1 - \mathbf{f}_i^1)\mathbf{v}_k \right],$$

respectively. The discrete first fundamental form of the mapping $f^{-1}$ is defined as

$$I_{f^{-1}} = \begin{bmatrix} \mathbf{g}_1^\top \mathbf{g}_1 & \mathbf{g}_1^\top \mathbf{g}_2 \\ \mathbf{g}_1^\top \mathbf{g}_2 & \mathbf{g}_2^\top \mathbf{g}_2 \end{bmatrix},$$

where $\mathbf{g}_1 = \frac{\partial f^{-1}}{\partial z^1}$ and $\mathbf{g}_2 = \frac{\partial f^{-1}}{\partial z^1}$. A bijective mapping $f : \mathcal{M} \to \mathbb{D}$ is said to be equiareal if

$$\det(I_{f^{-1}}) = 1.$$

## 3 Previous Works

In this section, we briefly review the fast stretch minimization (FSM) algorithm for the computation of an equiareal parameterization of a simply connected open mesh $\mathcal{M}$ proposed by Yoshizawa et al. [20], which is the most related work. The computational procedure of FSM [20] is described as follows. First, an initial mapping is given by the discrete harmonic mapping [10] described in Sect. 3.1. Next, an iterative scheme is applied to the FSM algorithm for improving the interior vertices to achieve an equiareal parameterization, which is described in Sect. 3.2.

### 3.1 Discrete Harmonic Mappings

Given a simply connected open mesh $\mathcal{M}$, the discrete harmonic mapping is the minimizer of the discrete Dirichlet energy functional defined as

$$E(\mathbf{f}) = \frac{1}{2} \sum_{[\mathbf{v}_i, \mathbf{v}_j] \in \mathcal{E}(\mathcal{M})} w_{i,j} |\mathbf{f}_i - \mathbf{f}_j|^2, \tag{1}$$

where $w_{i,j}$ is the cotangent weight on the edge $[\mathbf{v}_i, \mathbf{v}_j]$ defined by

$$w_{i,j} = \frac{\cot \alpha_{i,j} + \cot \alpha_{j,i}}{2}$$

in which $\alpha_{i,j}$ and $\alpha_{j,i}$ are two angles opposite to the edge $[\mathbf{v}_i, \mathbf{v}_j]$, as illustrated in Fig. 1. The energy functional (1) can be written as

$$E(\mathbf{f}) = \frac{1}{2} \mathbf{f}^* L \mathbf{f},$$

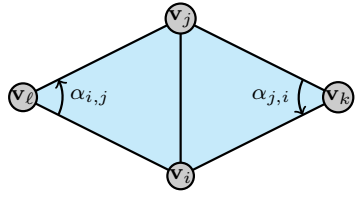

**Fig. 1** An illustration for the cotangent weights.

where $\mathbf{f}^*$ denotes the conjugate transpose of $\mathbf{f}$ and $L$ is the Laplacian matrix defined by

$$L_{i,j} = \begin{cases} -w_{i,j} & \text{if } [\mathbf{v}_i, \mathbf{v}_j] \in \mathcal{E}(\mathcal{M}), \\ \sum_{k \neq i} w_{i,k} & \text{if } j = i, \\ 0 & \text{otherwise.} \end{cases} \tag{2}$$

Suppose I and B are the sets of indices of the interior vertices and the boundary vertices, respectively. The discrete harmonic mapping can be computed by solving the discrete Laplace-Beltrami equation

$$L_{\mathtt{I},\mathtt{I}} \mathbf{f}_{\mathtt{I}} = -L_{\mathtt{I},\mathtt{B}} \mathbf{f}_{\mathtt{B}},$$

where $\mathbf{f}_{\mathtt{B}}$ is the boundary mapping given by

$$(\mathbf{f}_{\mathtt{B}})_j = \cos\theta_j + \mathrm{i}\sin\theta_j$$

in which

$$\theta_j = 2\pi \frac{\sum_{\ell=1}^{j} |(\mathbf{v}_{\mathtt{B}})_{\ell+1} - (\mathbf{v}_{\mathtt{B}})_\ell|}{\sum_{\ell=1}^{n_{\mathtt{B}}} |(\mathbf{v}_{\mathtt{B}})_{\ell+1} - (\mathbf{v}_{\mathtt{B}})_\ell|},$$

and $(\mathbf{v}_{\mathtt{B}})_{n_{\mathtt{B}}+1} := (\mathbf{v}_{\mathtt{B}})_1$, for $j = 1, \ldots, n_{\mathtt{B}}$.

3.2 Iterative Scheme of FSM Algorithm

Given a bijective piecewise affine mapping $f : \mathcal{M} \to \mathcal{D} \subset \mathbb{C}$. The singular values of the Jacobian matrix $J_{f^{-1}}(z) = \left[\frac{\partial f^{-1}}{\partial z^1}(z) \ \frac{\partial f^{-1}}{\partial z^2}(z)\right]$ are given by

$$\gamma_{\mathbf{f}}^{\pm}(z) = \sqrt{\frac{(a_{\mathbf{f}} + c_{\mathbf{f}}) \pm \sqrt{(a_{\mathbf{f}} - c_{\mathbf{f}})^2 + 4b_{\mathbf{f}}^2}}{2}},$$

where $a_{\mathbf{f}} = (\frac{\partial f^{-1}}{\partial z^1})^\top (\frac{\partial f^{-1}}{\partial z^1})$, $b_{\mathbf{f}} = (\frac{\partial f^{-1}}{\partial z^1})^\top (\frac{\partial f^{-1}}{\partial z^2})$, and $c_{\mathbf{f}} = (\frac{\partial f^{-1}}{\partial z^2})^\top (\frac{\partial f^{-1}}{\partial z^2})$. Note that $\gamma_{\mathbf{f}}^{\pm}|_\tau(z)$ are constants on each triangular face $\tau \in \mathcal{F}(\mathcal{M})$. The *stretch metric* over $\tau$ is defined as

$$\mathcal{S}_{\mathbf{f}}(\tau) = \sqrt{\frac{(\gamma_{\mathbf{f}}^+|_\tau)^2 + (\gamma_{\mathbf{f}}^-|_\tau)^2}{2}} = \sqrt{\frac{a_{\mathbf{f}}|_\tau + c_{\mathbf{f}}|_\tau}{2}}, \tag{3}$$

i.e., the root-mean-square of the singular values of $J_{f^{-1}}$ [16]. Then the stretch metric on the vertex $\mathbf{v}_j$ is defined as

$$\mathcal{T}_{\mathbf{f}^{(k)}}(\mathbf{v}_j) = \sqrt{\frac{\sum_{\tau \in \mathcal{N}(\mathbf{v}_j)} |\tau| \mathcal{S}_{\mathbf{f}^{(k)}}(\tau)^2}{\sum_{\tau \in \mathcal{N}(\mathbf{v}_j)} |\tau|}},$$

in which $\mathcal{N}(\mathbf{v}_j) = \{\tau \in \mathcal{F}(\mathcal{M}) \,|\, \mathbf{v}_j \in \tau\}$, $\mathcal{S}_{\mathbf{f}^{(k)}}(\tau)$ is the stretch metric at the $k$-th iterate as in (3), and $|\tau|$ denotes the area of the triangular face $\tau \in \mathcal{F}(\mathcal{M})$.

The interior vertices of the mapping is updated via redistributing the local stretches by assigning $w_{i,j}^{(0)} := w_{i,j}$ and

$$w_{i,j}^{(k+1)} = \frac{w_{i,j}^{(k)}}{\mathcal{T}_{\mathbf{f}^{(k)}}(\mathbf{v}_j)}.$$

In other words, if we let the weight matrix $W^{(k)}$ be

$$W_{i,j}^{(k)} = \begin{cases} w_{i,j}^{(k)} & \text{if } [\mathbf{v}_i, \mathbf{v}_j] \in \mathcal{E}(\mathcal{M}), \\ 0 & \text{otherwise,} \end{cases}$$

and define the diagonal matrix of the stretch metric on vertices as

$$D_{\mathbf{f}^{(k)}} = \mathrm{diag}\left(\left[\frac{1}{\mathcal{T}_{\mathbf{f}^{(k)}}(\mathbf{v}_j)}\right]_{j=1}^n\right),$$

then the interior vertices $\mathbf{f}_\mathtt{I}$ is updated by the iteration

$$L_{\mathtt{I},\mathtt{I}}^{(k)} \mathbf{f}_\mathtt{I}^{(k)} = -L_{\mathtt{I},\mathtt{B}}^{(k)} \mathbf{f}_\mathtt{B}, \tag{4}$$

where $\mathbf{f}_\mathtt{I}^{(0)} \equiv \mathbf{f}_\mathtt{I}$, $L^{(0)} \equiv L$ as in (2), and

$$L^{(k+1)} = \mathrm{diag}\left(\left[\sum_{j=1}^n [W^{(k)} D_{\mathbf{f}^{(k)}}]_{i,j}\right]_{i=1}^n\right) - W^{(k)} D_{\mathbf{f}^{(k)}}.$$

Note that the stretch metric (3) cannot distinguish some different types of the first fundamental forms, e.g.,

$$\begin{bmatrix} 1 & 0 \\ 0 & 1 \end{bmatrix}, \quad \begin{bmatrix} 1.2 & 0 \\ 0 & 0.8 \end{bmatrix}, \quad \text{and} \quad \begin{bmatrix} 1 & 0.3 \\ 0.2 & 1 \end{bmatrix}.$$

To overcome this issue, the stretch metric needs to be modified.

*Remark 1* Note that the matrix $L_{\mathtt{I},\mathtt{I}}^{(k)}$ in the linear system (4), in general, is not symmetric.

## 4 The Stretch Energy Minimization Algorithm

In this section, we describe the SEM algorithm for computing an equiareal parameterization $f$ of a simply connected open surface $\mathcal{M}$. First, the initial mapping is given by the harmonic mapping [10] in Sect. 3.1. Then an iterative scheme of SEM is proposed to achieve an equiareal parameterization, which will be introduced in Sect. 4.1.

4.1 Iterative Scheme of SEM Algorithm

Given a bijective piecewise affine mapping $f : \mathcal{M} \to \mathbb{D} \subset \mathbb{C}$. Write

$$\mathbf{f} = (f(\mathbf{v}_1), \ldots, f(\mathbf{v}_n))^\top \equiv (\mathbf{f}_1, \ldots, \mathbf{f}_n)^\top.$$

The local stretch factor $\sigma_{\mathbf{f}} : \mathcal{F}(\mathcal{M}) \to \mathbb{R}$ with respect to $\mathbf{f}$ is defined by

$$\sigma_{\mathbf{f}}([\mathbf{v}_i, \mathbf{v}_j, \mathbf{v}_k]) = \frac{|[\mathbf{v}_i, \mathbf{v}_j, \mathbf{v}_k]|}{|[\mathbf{f}_i, \mathbf{f}_j, \mathbf{f}_k]|}, \tag{5}$$

where $|[\mathbf{v}_i, \mathbf{v}_j, \mathbf{v}_k]|$ and $|[\mathbf{f}_i, \mathbf{f}_j, \mathbf{f}_k]|$ denote the areas of the triangular faces $[\mathbf{v}_i, \mathbf{v}_j, \mathbf{v}_k]$ and $[\mathbf{f}_i, \mathbf{f}_j, \mathbf{f}_k]$, respectively. The proof for the consistency of

$$\sigma_{\mathbf{f}}([\mathbf{v}_i, \mathbf{v}_j, \mathbf{v}_k])^2 = \det \left( I_{f^{-1}} |_{[\mathbf{f}_i, \mathbf{f}_j, \mathbf{f}_k]} \right)$$

can be found in Appendix A. The stretched Laplacian matrix $L(\mathbf{f})$ is defined as

$$[L(\mathbf{f})]_{i,j} = \begin{cases} -\omega_{i,j}(\mathbf{f}) & \text{if } [\mathbf{v}_i, \mathbf{v}_j] \in \mathcal{E}(\mathcal{M}), \\ \sum_{\ell \neq i} \omega_{i,\ell}(\mathbf{f}) & \text{if } j = i, \\ 0 & \text{otherwise}, \end{cases} \tag{6}$$

where $\omega_{i,j}(\mathbf{f})$ is the stretched cotangent weight given by

$$\omega_{i,j}(\mathbf{f}) = \frac{1}{2} \left( \frac{\cot(\alpha_{i,j}(\mathbf{f}))}{\sigma_{\mathbf{f}}([\mathbf{v}_i, \mathbf{v}_j, \mathbf{v}_k])} + \frac{\cot(\alpha_{j,i}(\mathbf{f}))}{\sigma_{\mathbf{f}}([\mathbf{v}_j, \mathbf{v}_i, \mathbf{v}_\ell])} \right)$$

in which $\alpha_{i,j}(\mathbf{f})$ and $\alpha_{j,i}(\mathbf{f})$ are two angles opposite to the edge $[\mathbf{f}_i, \mathbf{f}_j]$ connecting vertices $\mathbf{f}_i$ and $\mathbf{f}_j$ on $\mathbb{C}$. We define the *stretch energy* as

$$E_S(\mathbf{f}) = \frac{1}{2} \mathbf{f}^* L(\mathbf{f}) \mathbf{f}, \tag{7}$$

where $L(\mathbf{f})$ is the stretched Laplacian matrix as in (6).

*Remark 2* Note that

$$\cot(\alpha_{i,j}(\mathbf{f})) = \frac{(\mathbf{f}_i^1 - \mathbf{f}_k^1)(\mathbf{f}_j^1 - \mathbf{f}_k^1) + (\mathbf{f}_i^2 - \mathbf{f}_k^2)(\mathbf{f}_j^2 - \mathbf{f}_k^2)}{|(\mathbf{f}_i^1 - \mathbf{f}_k^1)(\mathbf{f}_j^2 - \mathbf{f}_k^2) - (\mathbf{f}_i^2 - \mathbf{f}_k^2)(\mathbf{f}_j^1 - \mathbf{f}_k^1)|}$$
$$= \frac{\sum_{s=1}^{2}(\mathbf{f}_i^s - \mathbf{f}_k^s)(\mathbf{f}_j^s - \mathbf{f}_k^s)}{2|[\mathbf{f}_i, \mathbf{f}_j, \mathbf{f}_k]|}.$$

The ratio between $\cot(\alpha_{i,j}(\mathbf{f}))$ and $\sigma_{\mathbf{f}}([\mathbf{v}_i, \mathbf{v}_j, \mathbf{v}_k])$ can be simplified as

$$\frac{\cot(\alpha_{i,j}(\mathbf{f}))}{\sigma_{\mathbf{f}}([\mathbf{v}_i, \mathbf{v}_j, \mathbf{v}_k])} = \frac{\sum_{s=1}^{2}(\mathbf{f}_i^s - \mathbf{f}_k^s)(\mathbf{f}_j^s - \mathbf{f}_k^s)}{2|[\mathbf{v}_i, \mathbf{v}_j, \mathbf{v}_k]|}.$$

As a result, the stretched cotangent weight $\omega_{i,j}(\mathbf{f})$ can be written as

$$\omega_{i,j}(\mathbf{f}) = \frac{1}{2} \sum_{[\mathbf{v}_i, \mathbf{v}_j, \mathbf{v}_k] \in \mathcal{F}(\mathcal{M})} \sum_{s=1}^{2} \frac{(\mathbf{f}_i^s - \mathbf{f}_k^s)(\mathbf{f}_j^s - \mathbf{f}_k^s)}{2|[\mathbf{v}_i, \mathbf{v}_j, \mathbf{v}_k]|}.$$

Our aim is to find a minimizer for the stretch energy functional (7). Note that $\mathbf{f}_\ell = \mathbf{f}_\ell^1 + i\mathbf{f}_\ell^2$ in which $\mathbf{f}_\ell^1, \mathbf{f}_\ell^2 \in \mathbb{R}$, for $k = 1, \ldots, n$. The stretch energy (7) can be written as

$$E_S(\mathbf{f}^1, \mathbf{f}^2) = \frac{1}{2}\left((\mathbf{f}^1)^\top L(\mathbf{f})\mathbf{f}^1 + (\mathbf{f}^2)^\top L(\mathbf{f})\mathbf{f}^2\right).$$

Then the gradient

$$\nabla E_S(\mathbf{f}^1, \mathbf{f}^2) = \left(\frac{\partial E_S(\mathbf{f}^1, \mathbf{f}^2)}{\partial \mathbf{f}^1}, \frac{\partial E_S(\mathbf{f}^1, \mathbf{f}^2)}{\partial \mathbf{f}^2}\right)$$

of the energy functional can be calculated by

$$\frac{\partial E_S(\mathbf{f}^1, \mathbf{f}^2)}{\partial \mathbf{f}^s} = L(\mathbf{f})\mathbf{f}^s + \frac{1}{2}\sum_{t=1}^{2}(\mathbf{f}^t)^\top \frac{\partial L(\mathbf{f})}{\partial \mathbf{f}^s}\mathbf{f}^t, \tag{8}$$

where $\frac{\partial L(\mathbf{f})}{\partial \mathbf{f}^s}$ is the tensor with entries as

$$\frac{\partial [L(\mathbf{f})]_{i,j}}{\partial \mathbf{f}_k^s} = \begin{cases} -\dfrac{1}{2}\displaystyle\sum_{[\mathbf{v}_i,\mathbf{v}_j,\mathbf{v}_\ell]\in\mathcal{F}(\mathcal{M})}\dfrac{\mathbf{f}_j^s - \mathbf{f}_\ell^s}{2|[\mathbf{v}_i,\mathbf{v}_j,\mathbf{v}_\ell]|} & \text{if } [\mathbf{v}_i,\mathbf{v}_j]\in\mathcal{E}(\mathcal{M}),\ k=i,\\ -\dfrac{1}{2}\displaystyle\sum_{[\mathbf{v}_i,\mathbf{v}_j,\mathbf{v}_\ell]\in\mathcal{F}(\mathcal{M})}\dfrac{\mathbf{f}_i^s - \mathbf{f}_\ell^s}{2|[\mathbf{v}_i,\mathbf{v}_j,\mathbf{v}_\ell]|} & \text{if } [\mathbf{v}_i,\mathbf{v}_j]\in\mathcal{E}(\mathcal{M}),\ k=j,\\ -\dfrac{1}{2}\dfrac{2\mathbf{f}_k^s - \mathbf{f}_i^s - \mathbf{f}_j^s}{2|[\mathbf{v}_i,\mathbf{v}_j,\mathbf{v}_k]|} & \text{if } [\mathbf{v}_i,\mathbf{v}_j,\mathbf{v}_k]\in\mathcal{F}(\mathcal{M}),\\ 0 & \text{otherwise,} \end{cases}$$

for $s = 1, 2$. A direct computation yields that

$$\left[(\mathbf{f}^s)^\top \frac{\partial L(\mathbf{f})}{\partial \mathbf{f}^s}\right]_{j,k} = \begin{cases} -\dfrac{1}{2}\displaystyle\sum_{[\mathbf{v}_\alpha,\mathbf{v}_j,\mathbf{v}_k]\in\mathcal{F}(\mathcal{M})}\dfrac{(\mathbf{f}_k^s - \mathbf{f}_\alpha^s)(\mathbf{f}_\alpha^s - \mathbf{f}_j^s)}{2|[\mathbf{v}_\alpha,\mathbf{v}_j,\mathbf{v}_k]|} & \text{if } [\mathbf{v}_j,\mathbf{v}_k]\in\mathcal{E}(\mathcal{M}),\\ -\dfrac{1}{2}\displaystyle\sum_{[\mathbf{v}_\alpha,\mathbf{v}_j,\mathbf{v}_\beta]\in\mathcal{F}(\mathcal{M})}\dfrac{(\mathbf{f}_\alpha^s - \mathbf{f}_\beta^s)^2}{2|[\mathbf{v}_\alpha,\mathbf{v}_j,\mathbf{v}_\beta]|} & \text{if } k = j,\\ 0 & \text{otherwise,} \end{cases} \tag{9}$$

for $s = 1, 2$. Similarly,

$$\left[(\mathbf{f}^s)^\top \frac{\partial L(\mathbf{f})}{\partial \mathbf{f}^t}\right]_{j,k}$$
$$= \begin{cases} -\dfrac{1}{2}\displaystyle\sum_{[\mathbf{v}_\alpha,\mathbf{v}_j,\mathbf{v}_k]\in\mathcal{F}(\mathcal{M})}\dfrac{2(\mathbf{f}_k^s - \mathbf{f}_\alpha^s)(\mathbf{f}_\alpha^t - \mathbf{f}_j^t) - (\mathbf{f}_j^s - \mathbf{f}_\alpha^s)(\mathbf{f}_\alpha^t - \mathbf{f}_k^t)}{2|[\mathbf{v}_\alpha,\mathbf{v}_j,\mathbf{v}_k]|} & \text{if } [\mathbf{v}_j,\mathbf{v}_k]\in\mathcal{E}(\mathcal{M}),\\ -\dfrac{1}{2}\displaystyle\sum_{[\mathbf{v}_\alpha,\mathbf{v}_j,\mathbf{v}_\beta]\in\mathcal{F}(\mathcal{M})}\dfrac{(\mathbf{f}_\alpha^s - \mathbf{f}_\beta^s)(\mathbf{f}_\alpha^t - \mathbf{f}_\beta^t)}{2|[\mathbf{v}_\alpha,\mathbf{v}_j,\mathbf{v}_\beta]|} & \text{if } k = j,\\ 0 & \text{otherwise,} \end{cases} \tag{10}$$

for $(s, t) = (1, 2), (2, 1)$. The derivations of (9) and (10) can be found in Appendix B for details.

In practice, we omit the quadratic terms in (8) and propose the iterative scheme for equiareal parameterizations as follows. First, the mapping of boundary vertices is updated by solving the linear system

$$\left[L(\mathbf{f}^{(k)})\right]_{\mathtt{B},\mathtt{B}} \mathbf{f}_{\mathtt{B}}^{(k+1)} = -\left[L(\mathbf{f}^{(k)})\right]_{\mathtt{B},\mathtt{I}} \mathbf{f}_{\mathtt{I}}^{(k)}. \qquad (11)$$

To ensure that the boundary vertices are mapped on the unit circle, we centralize the mapping by

$$\mathbf{f}_{\mathtt{B}}^{(k+1)} \leftarrow \left(I_{n_{\mathtt{B}}} - \frac{\mathbf{1}_{n_{\mathtt{B}}}\mathbf{1}_{n_{\mathtt{B}}}^{\top}}{n_{\mathtt{B}}}\right) \mathbf{f}_{\mathtt{B}}^{(k+1)}, \qquad (12)$$

and normalize the mapping by

$$\mathbf{f}_{\mathtt{B}}^{(k+1)} \leftarrow \mathrm{diag}\left(\left|\mathbf{f}_{\mathtt{B}}^{(k+1)}\right|^{-1}\right) \mathbf{f}_{\mathtt{B}}^{(k+1)}. \qquad (13)$$

Finally, the mapping of interior vertices is updated by solving the linear system

$$\left[L(\mathbf{f}^{(k)})\right]_{\mathtt{I},\mathtt{I}} \mathbf{f}_{\mathtt{I}}^{(k+1)} = -\left[L(\mathbf{f}^{(k)})\right]_{\mathtt{I},\mathtt{B}} \mathbf{f}_{\mathtt{B}}^{(k+1)}. \qquad (14)$$

Equivalently, the iterations (11)-(14) can be written as

$$\mathbf{f}_{\mathtt{B}}^{(k+1)} = N^{(k)} C K^{(k)} \mathbf{f}_{\mathtt{B}}^{(k)}, \qquad (15)$$

where

$$K^{(0)} = \left[L(\mathbf{f}^{(0)})\right]_{\mathtt{B},\mathtt{B}}^{-1} \left[L(\mathbf{f}^{(0)})\right]_{\mathtt{B},\mathtt{I}} L_{\mathtt{I},\mathtt{I}}^{-1} L_{\mathtt{I},\mathtt{B}},$$

and

$$K^{(k)} = \left[L(\mathbf{f}^{(k)})\right]_{\mathtt{B},\mathtt{B}}^{-1} \left[L(\mathbf{f}^{(k)})\right]_{\mathtt{B},\mathtt{I}} \left[L(\mathbf{f}^{(k-1)})\right]_{\mathtt{I},\mathtt{I}}^{-1} \left[L(\mathbf{f}^{(k-1)})\right]_{\mathtt{I},\mathtt{B}},$$

for $k \in \mathbb{N}$, $C$ and $N^{(k)}$ are the centralization and normalization matrices, respectively, given by

$$C = I_{n_{\mathtt{B}}} - \frac{\mathbf{1}_{n_{\mathtt{B}}}\mathbf{1}_{n_{\mathtt{B}}}^{\top}}{n_{\mathtt{B}}}$$

and

$$N^{(k)} = \mathrm{diag}\left(\left|C\left[L(\mathbf{f}^{(k)})\right]_{\mathtt{B},\mathtt{B}}^{-1} \left[L(\mathbf{f}^{(k)})\right]_{\mathtt{B},\mathtt{I}} \mathbf{f}_{\mathtt{I}}^{(k)}\right|^{-1}\right).$$

The SEM algorithm for equiareal parameterizations is summarized in Algorithm 1.

*Remark 3* The matrices $[L(\mathbf{f}^{(k)})]_{\mathtt{B},\mathtt{B}}$ and $[L(\mathbf{f}^{(k)})]_{\mathtt{I},\mathtt{I}}$ in (11) and (14), respectively, are symmetric positive definite so that the Cholesky decompositions are applicable in solving linear systems in steps 4, 8 and 11 in Algorithm 1.

*Remark 4* Several experiments indicate that the resulting equiareal parameterization would be slightly improved by performing an inversion transformation

$$(\mathbf{f}_{\mathtt{I}}^{(k)})_\ell \leftarrow \frac{1}{(\mathbf{f}_{\mathtt{I}}^{(k)})_\ell^*},$$

for $\ell = 1, \ldots, n_{\mathtt{I}}$, before solving the linear system (11). In other words, Eq. (11) is replaced by the equation

$$\left[L(\mathbf{f}^{(k)})\right]_{\mathtt{B},\mathtt{B}} \mathbf{f}_{\mathtt{B}}^{(k+1)} = -\left[L(\mathbf{f}^{(k)})\right]_{\mathtt{B},\mathtt{I}} \mathrm{diag}\left(\left|\mathbf{f}_{\mathtt{I}}^{(k)}\right|^{-2}\right) \mathbf{f}_{\mathtt{I}}^{(k)}.$$

---

**Algorithm 1** Stretch Energy Minimization (SEM)

**Input:** A mesh $\mathcal{M}$ of a simply connected open surface.
**Output:** An equiareal parameterization $\mathbf{f}$.
1: Compute the mesh area $\mathcal{A}(\mathcal{M}) := \sum_{\tau \in \mathcal{F}(\mathcal{M})} |\tau|$.
2: Scale the mesh $\mathcal{M} \leftarrow \frac{\pi}{\mathcal{A}(\mathcal{M})} \mathcal{M}$.
3: Compute the initial boundary mapping $\mathbf{f}_\texttt{B}$ by

$$(\mathbf{f}_\texttt{B})_j = \cos\theta_j + \mathrm{i}\sin\theta_j,$$

in which

$$\theta_j = 2\pi \frac{\sum_{\ell=1}^{j} |(\mathbf{v}_\texttt{B})_{\ell+1} - (\mathbf{v}_\texttt{B})_\ell|}{\sum_{\ell=1}^{n_\texttt{B}} |(\mathbf{v}_\texttt{B})_{\ell+1} - (\mathbf{v}_\texttt{B})_\ell|},$$

for $j = 1, \ldots, n_\texttt{B}$, $(\mathbf{v}_\texttt{B})_{n_\texttt{B}+1} := (\mathbf{v}_\texttt{B})_1$.
4: Initial interior mapping via $L_{\texttt{I},\texttt{I}} \mathbf{f}_\texttt{I} = -L_{\texttt{I},\texttt{B}} \mathbf{f}_\texttt{B}$.
5: Compute the stretching factor $\sigma_\mathbf{f}$.
6: Update the Laplacian matrix $L \leftarrow L(\mathbf{f})$.
7: **while** not convergent **do**
8:     Update boundary $L_{\texttt{B},\texttt{B}} \mathbf{f}_\texttt{B} = -L_{\texttt{B},\texttt{I}} \mathbf{f}_\texttt{I}$.
9:     Centralize boundary $\mathbf{f}_\texttt{B} \leftarrow \left( I_{n_\texttt{B}} - \frac{\mathbf{1}_{n_\texttt{B}} \mathbf{1}_{n_\texttt{B}}^\top}{n_\texttt{B}} \right) \mathbf{f}_\texttt{B}$.
10:    Normalize boundary $\mathbf{f}_\texttt{B} \leftarrow \mathrm{diag}\left( |\mathbf{f}_\texttt{B}|^{-1} \right) \mathbf{f}_\texttt{B}$.
11:    Update interior $L_{\texttt{I},\texttt{I}} \mathbf{f}_\texttt{I} = -L_{\texttt{I},\texttt{B}} \mathbf{f}_\texttt{B}$.
12:    Update the stretching factor $\sigma_\mathbf{f}$.
13:    Update the Laplacian matrix $L \leftarrow L(\mathbf{f})$.
14: **end while**

---

## 5 Existence of Nontrivial Limit Points

In this section, we aim to prove the existence of a nontrivial (nonconstant) limit point of the iteration of the proposed SEM algorithm. For convenience, we give a mild assumption for the triangular mesh.

**Definition 1 (Well-conditioned mesh)** A simply connected open mesh $\mathcal{M}$ is said to be *well-conditioned* if it satisfies the following conditions:

(i) The subgraph of all the interior vertices is connected.
(ii) Every boundary vertex is connected to at least one interior vertex.

Also, we give the definition of M-matrix [7] and some related lemmas.

**Definition 2** (i) A matrix $A \in \mathbb{R}^{m \times n}$ is said to be nonnegative (positive) if all the entries of $A$ are nonnegative (positive).
(ii) A squared matrix $A \in \mathbb{R}^{n \times n}$ is irreducible, if the corresponding graph $\mathcal{G}(A)$ of $A$ is connected.

**Definition 3** A matrix $A \in \mathbb{R}^{n \times n}$ is said to be an M-matrix if $A = sI - B$, where $B$ is nonnegative and $s \geq \rho(B)$, where $\rho(B)$ is the spectral radius of $B$.

**Lemma 1 (Theorem 1.4.10 in [15])** *Suppose $A \in \mathbb{R}^{n \times n}$ is a singular, irreducible M-matrix. Then each principal submatrix of $A$ other than $A$ itself is a nonsingular M-matrix.*

**Lemma 2 (Theorem 1.4.7 in [15])** *If $A \in \mathbb{R}^{n \times n}$ is a nonsingular M-matrix, then $A^{-1}$ is a nonnegative matrix. Moreover, if $A$ is irreducible, then $A^{-1}$ is a positive matrix.*

**Lemma 3 (Perron Theorem [14])** *Let $A \in \mathbb{R}^{n \times n}$ be a positive matrix. Then*

*(i) $\rho(A)$ is an eigenvalue of $A$, and all the other eigenvalues are strictly smaller than $\rho(A)$ in modulus.*

*(ii) $\rho(A)$ is the only eigenvalue that has a positive eigenvector.*

*(iii) $\rho(A)$ has algebraic multiplicity one.*

The following theorem plays an important role in the geometric point of view of the matrix product of $K^{(k)}$ in Eq. (15).

**Theorem 1** *Given a well-conditioned simply connected open mesh $\mathcal{M}$ of $n$ vertices. Let $L^{(1)}$ and $L^{(2)}$ be two Laplacian matrices of $\mathcal{M}$, defined similar as in (2), with positive weights $\{w_{i,j}^{(1)} \,|\, [\mathbf{v}_i, \mathbf{v}_j] \in \mathcal{E}(\mathcal{M})\}$ and $\{w_{i,j}^{(2)} \,|\, [\mathbf{v}_i, \mathbf{v}_j] \in \mathcal{E}(\mathcal{M})\}$, respectively. Let $\mathtt{I}$ and $\mathtt{B}$ be the index sets of interior vertices and boundary vertices, respectively. Let*
$$K = (L_{\mathtt{B},\mathtt{B}}^{(2)})^{-1} (L_{\mathtt{I},\mathtt{B}}^{(2)})^\top (L_{\mathtt{I},\mathtt{I}}^{(1)})^{-1} L_{\mathtt{I},\mathtt{B}}^{(1)}.$$
*Then $K$ is positive with $\rho(K) = 1$ being the unique largest eigenvalue of $K$ in modulus.*

*Proof* From the definition of the Laplacian matrix (2), it is clear that $L^{(t)} \mathbf{1}_{n_L} = \mathbf{0}$, for $t = 1, 2$. That is, for $t = 1, 2$,
$$\begin{cases} L_{\mathtt{I},\mathtt{I}}^{(t)} \mathbf{1}_{n_\mathtt{I}} + L_{\mathtt{I},\mathtt{B}}^{(t)} \mathbf{1}_{n_\mathtt{B}} = \mathbf{0}, \\ (L_{\mathtt{I},\mathtt{B}}^{(t)})^\top \mathbf{1}_{n_\mathtt{I}} + L_{\mathtt{B},\mathtt{B}}^{(t)} \mathbf{1}_{n_\mathtt{B}} = \mathbf{0}. \end{cases} \tag{16}$$

Note that $L^{(t)}$ is a singular irreducible M-matrix, for $t = 1, 2$. By Lemma 1, the matrices $L_{\mathtt{I},\mathtt{I}}^{(t)}$ and $L_{\mathtt{B},\mathtt{B}}^{(t)}$ are invertible, for $t = 1, 2$. Then Eq. (16) implies that
$$\begin{cases} (L_{\mathtt{I},\mathtt{I}}^{(t)})^{-1} L_{\mathtt{I},\mathtt{B}}^{(t)} \mathbf{1}_{n_\mathtt{B}} = -\mathbf{1}_{n_\mathtt{I}}, \\ (L_{\mathtt{B},\mathtt{B}}^{(t)})^{-1} (L_{\mathtt{I},\mathtt{B}}^{(t)})^\top \mathbf{1}_{n_\mathtt{I}} = -\mathbf{1}_{n_\mathtt{B}}. \end{cases} \tag{17}$$

It follows from Eq. (17) that
$$\begin{aligned} K \mathbf{1}_{n_\mathtt{B}} &= (L_{\mathtt{B},\mathtt{B}}^{(2)})^{-1} (L_{\mathtt{I},\mathtt{B}}^{(2)})^\top (L_{\mathtt{I},\mathtt{I}}^{(1)})^{-1} L_{\mathtt{I},\mathtt{B}}^{(1)} \mathbf{1}_{n_\mathtt{B}} \\ &= -(L_{\mathtt{B},\mathtt{B}}^{(2)})^{-1} (L_{\mathtt{I},\mathtt{B}}^{(2)})^\top \mathbf{1}_{n_\mathtt{I}} = \mathbf{1}_{n_\mathtt{B}}. \end{aligned}$$

That is, 1 is an eigenvalue of $K$ associated with the eigenvector $\mathbf{1}_{n_\mathtt{B}}$. On the other hand, the irreducibilities of $L_{\mathtt{I},\mathtt{I}}^{(1)}$ and $L_{\mathtt{B},\mathtt{B}}^{(2)}$ are respectively guaranteed by Definition 1 (i) and the assumption that $\mathcal{M}$ has only one boundary. By Lemma 2, $(L_{\mathtt{I},\mathtt{I}}^{(1)})^{-1}$ and $(L_{\mathtt{B},\mathtt{B}}^{(2)})^{-1}$ are positive. Furthermore, Definition 1 (ii) and the positivity of $(L_{\mathtt{I},\mathtt{I}}^{(1)})^{-1}$ guarantee that $(L_{\mathtt{I},\mathtt{B}}^{(2)})^\top (L_{\mathtt{I},\mathtt{I}}^{(1)})^{-1} L_{\mathtt{I},\mathtt{B}}^{(1)}$ is positive. Then the positivity of $K$ is guaranteed by the positivity of $(L_{\mathtt{B},\mathtt{B}}^{(2)})^{-1}$. Hence, By Lemma 3, $\rho(K) = 1$ is the largest eigenvalue of $K$ with algebraic multiplicity one, and all the other eigenvalues are strictly smaller than 1 in modulus. □

*Remark 5* Geometrically, for $i = 1, \ldots, n_K$,
$$(K\mathbf{f})_i = \sum_{j=1}^{n_K} K_{i,j} \mathbf{f}_j$$

is a strictly convex combination of the points $\{\mathbf{f}_j \in \mathbb{C}\}_{j=1}^{n_K}$.

*Remark 6* In general, if some weights of the stretched Laplacian matrix (6) are negative, the flip algorithm [8] can be applied to achieve positive weights.

In the following, we prove the existence of limit points of the iterative scheme (15).

**Theorem 2** *The sequence $\{\mathbf{f}_\mathtt{B}^{(k)}\}_{k\in\mathbb{N}}$ defined by (15) has a limit point $\mathbf{f}_\mathtt{B}^{(*)} \neq \mathbf{1}_{n_\mathtt{B}}$.*

*Proof* Since every entry of $\mathbf{f}_\mathtt{B}^{(k)}$ is on the unit circle, by Bolzano-Weierstrass theorem there exists a vector $\mathbf{f}_\mathtt{B}^{(*)}$ and a convergent subsequence $\{\mathbf{f}_\mathtt{B}^{(k_j)}\}_{j\in\mathbb{N}}$ such that

$$\lim_{j\to\infty} \mathbf{f}_\mathtt{B}^{(k_j)} = \mathbf{f}_\mathtt{B}^{(*)}.$$

From Remark 5, the centralization in the iteration (15) guarantees that after a rotation by setting $(\mathbf{f}_\mathtt{B}^{(k)})_1 = 1$ for each $k \in \mathbb{N}$, the maximal argument satisfies

$$\max_{1\leq\ell\leq n_\mathtt{B}} \mathrm{Arg}\left((CK^{(k)}\mathbf{f}_\mathtt{B}^{(k)})_\ell\right) > \pi. \tag{18}$$

Otherwise, each entry of the vector $CK^{(k)}\mathbf{f}_\mathtt{B}^{(k)}$ is located on the upper half-plane of $\mathbb{C}$. Then the center

$$\frac{1}{n_\mathtt{B}} \sum_{i=1}^{n_\mathtt{B}} (CK^{(k)}\mathbf{f}_\mathtt{B}^{(k)})_i \neq 0,$$

which contradicts that the center should be zero. In particular, Eq. (18) holds for the subsequence $\{k_j\}_{j\in\mathbb{N}}$. Hence, the accumulation point $\mathbf{f}_\mathtt{B}^{(*)}$ satisfies

$$\max_{1\leq\ell\leq n_\mathtt{B}} \mathrm{Arg}\left((CK^{(k)}\mathbf{f}_\mathtt{B}^{(*)})_\ell\right) \geq \pi.$$

Therefore, $\mathbf{f}_\mathtt{B}^{(*)} \neq \mathbf{1}_{n_\mathtt{B}}$. □

## 6 Numerical Experiments

In this section, we demonstrate the numerical results of the equiareal parameterizations obtained by SEM algorithm. The maximal numbers of iterations for SEM algorithm are set to be 10. The linear systems in SEM algorithm are solved using the backslash operator (\) in MATLAB.

Although the Lie advection method [24] based on Lie derivative and Cartans formula can handle surfaces with complicated topologies, however, the iterative procedure for optimizing triangulations in [24] is indirect and less efficient. Therefore, we compare the efficiency and the accuracy of our SEM algorithm to two of the state-of-the-art algorithms of equiareal parameterizations for simply connected open surfaces, namely, the fast stretch minimization (FSM) algorithm [20] and the optimal mass transportation (OMT) algorithm [18]. The MATLAB code of the FSM algorithm is reproduced by authors. The executable program files of the OMT algorithm are obtained from Gu's website [2]. Some of the mesh models are obtained from TurboSquid [5], AIM@SHAPE shape repository [3], the Stanford 3D scanning repository [4], and a project page of ALICE [1].

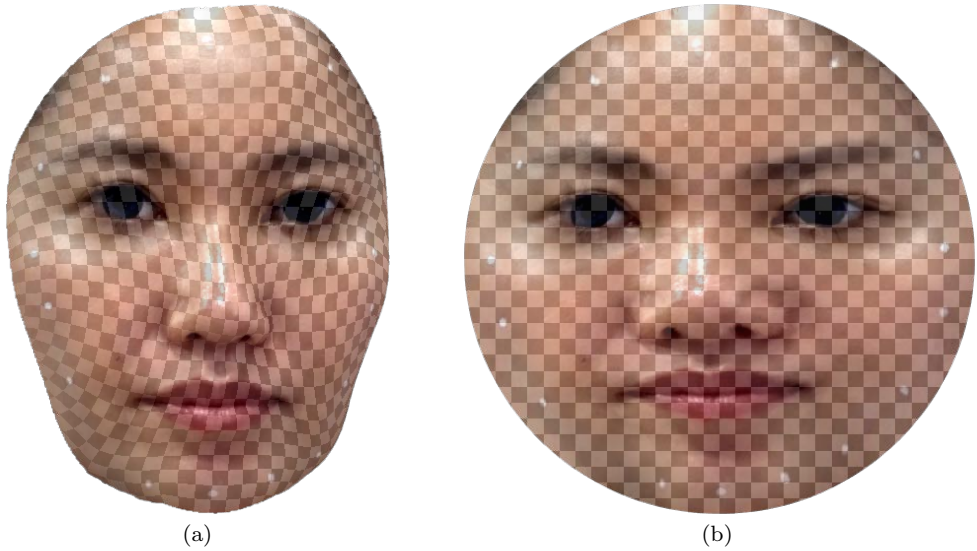

**Fig. 2** (a) The model of *Liu's Neutral Face*. (b) The equiareal parameterization obtained by SEM algorithm.

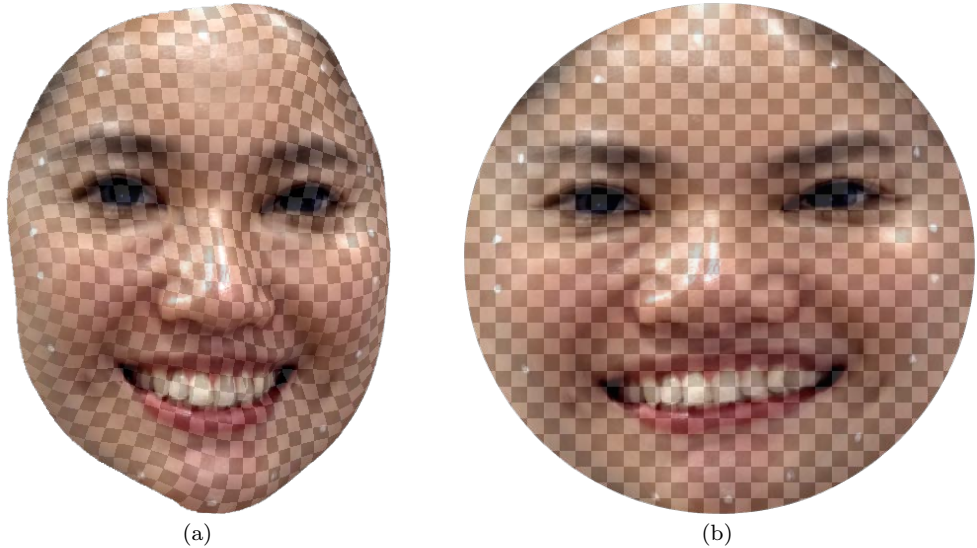

**Fig. 3** (a) The model of *Liu's Smiling Face*. (b) The equiareal parameterization obtained by SEM algorithm.

Figs. 2, 3, 4, and 5, respectively, show the models *Liu's Neutral Face*, *Liu's Smiling Face*, *Liu's Wry Face*, and *Liu's Pouting Face* and their equiareal parameterizations computed by the SEM algorithm.

A comparison of the computational cost between FSM, OMT and SEM algorithms is demonstrated in Table 1. The notation "−" means the iteration of the algorithm diverges. Also, Fig. 6 illustrates the relationship between the number of faces and the computational cost. As shown in Fig. 6 (a), the efficiency of SEM

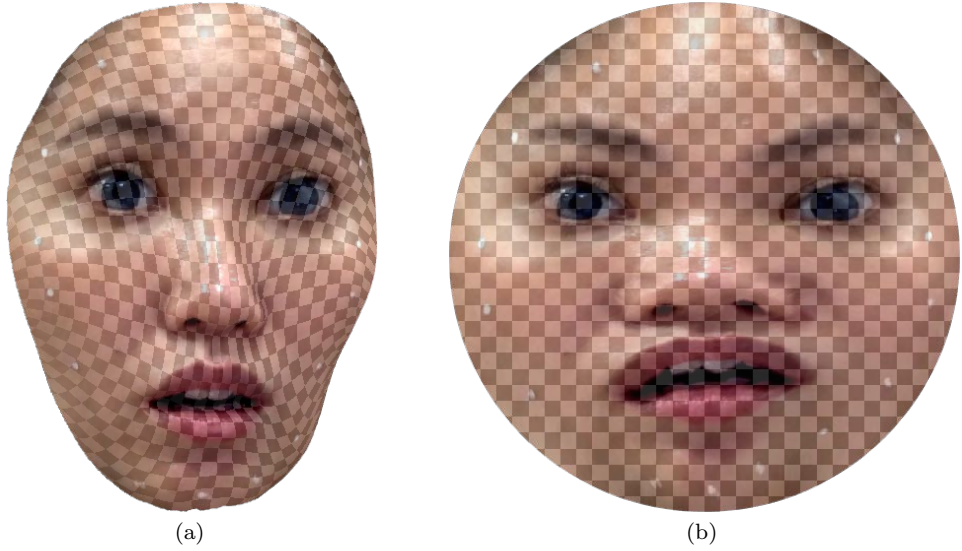

(a) (b)

**Fig. 4** (a) The model of *Liu's Wry Face*. (b) The equiareal parameterization obtained by SEM algorithm.

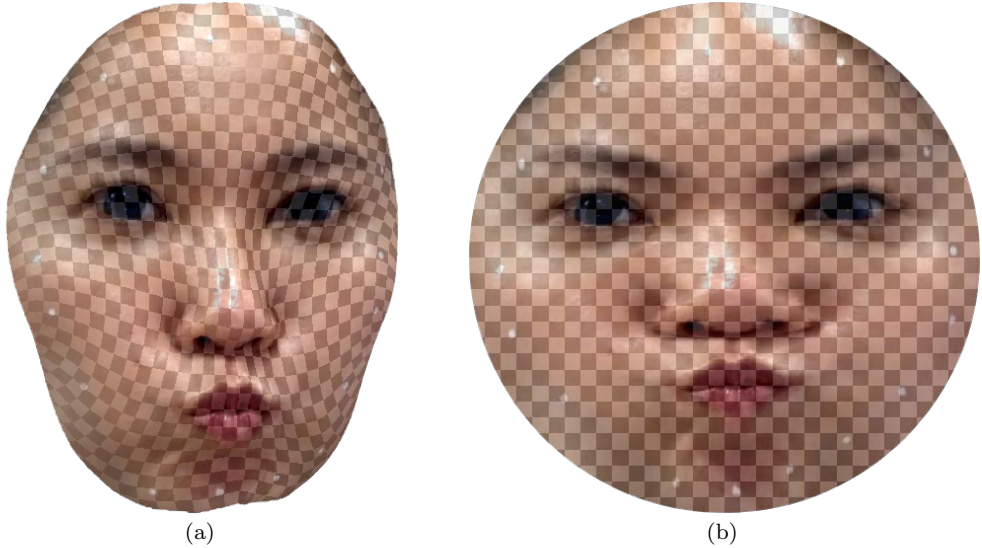

(a) (b)

**Fig. 5** (a) The model of *Liu's Pouting Face*. (b) The equiareal parameterization obtained by SEM algorithm.

outperforms the OMT, especially when the number of faces is large. In addition, as shown in Fig. 6 (b), the computational cost of FSM is roughly double of the computational cost of the proposed SEM algorithm. These results are consistent with the fact that the LU decomposition involves roughly $\frac{2}{3}n^3$ floating-point operations (FLOPs) while the Cholesky decomposition involves $\frac{1}{3}n^3$ FLOPs.

To measure the equiareal distortion, the *total area distortion* and the *local area ratio* for the parameterization mappings are commonly used. The total area

**Table 1** The computational cost (second) of the equiareal parameterizations by the FSM [20], OMT [18], and SEM algorithms. The notation "−" means the iteration of the algorithm diverges.

| Model Name | # Faces | FSM [20] Time | FSM [20] #Iter. | OMT [18] Time | OMT [18] #Iter. | SEM Time | SEM #Iter. |
|---|---|---|---|---|---|---|---|
| Nefertiti | 562 | 0.10 | 10 | 0.12 | 26 | 0.10 | 10 |
| Cowboy Hat | 4,604 | 0.14 | 10 | 1.37 | 68 | 0.08 | 10 |
| Chinese Vase | 5,592 | 0.14 | 10 | − | − | 0.07 | 10 |
| Bourbon Bottle | 13,088 | − | − | − | − | 0.23 | 10 |
| Foot | 19,966 | − | − | 7.11 | 104 | 0.36 | 10 |
| Chinese Lion | 34,421 | 1.24 | 10 | 4.29 | 26 | 0.69 | 10 |
| Stanford Bunny | 65,221 | − | − | − | − | 1.76 | 10 |
| Human Brain | 96,811 | − | − | 30.89 | 58 | 2.58 | 10 |
| Left Hand | 105,860 | − | − | − | − | 2.65 | 10 |
| Statue of Liberty | 190,162 | − | − | − | − | 5.04 | 10 |
| Liu's Neutral Face | 193,298 | 9.63 | 10 | 28.15 | 20 | 5.38 | 10 |
| Liu's Smiling Face | 205,207 | 10.22 | 10 | 30.56 | 20 | 5.65 | 10 |
| Liu's Pouting Face | 207,721 | 10.37 | 10 | 30.89 | 20 | 5.81 | 10 |
| Liu's Wry Face | 208,283 | 10.39 | 10 | 30.68 | 20 | 5.87 | 10 |
| Isis Statue | 374,309 | − | − | − | − | 12.52 | 10 |
| Bimba Statue | 836,740 | − | − | 859.29 | 124 | 28.16 | 10 |
| Knit Cap Man | 1,287,579 | − | − | 1753.11 | 66 | 70.23 | 10 |

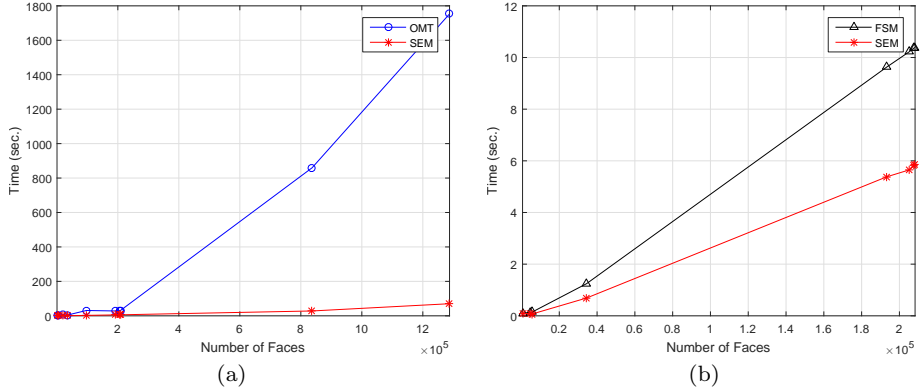

**Fig. 6** Computational cost (second) versus number of faces by FSM, OMT, and SEM.

distortion of a mapping $f$ on a mesh $\mathcal{M}$ is defined as

$$\mathcal{D}_\mathcal{M}(f) = \sum_{\tau \in \mathcal{F}(\mathcal{M})} \left| \frac{|\tau|}{\sum_{\tau \in \mathcal{F}(\mathcal{M})} |\tau|} - \frac{|f(\tau)|}{\sum_{\tau \in \mathcal{F}(\mathcal{M})} |f(\tau)|} \right|, \qquad (19)$$

where $|\tau|$ denotes the area of the triangular face $\tau$. A mapping $f$, defined on $\mathcal{M}$, is area-preserving if $\mathcal{D}_\mathcal{M}(f) = 0$. The local area distortion is measured using the mean and the standard deviation (SD) of the area ratio between the triangular faces on the surface $\mathcal{M}$ and on the image of the parameterization mapping $f$. In other words, for each triangular face $\tau \in \mathcal{F}(\mathcal{M})$, the local area ratio $\mathcal{R}_f : \mathcal{F}(\mathcal{M}) \to \mathbb{R}$ is defined as

$$\mathcal{R}_f(\tau) = \frac{|\tau| / \sum_{\tau \in \mathcal{F}(\mathcal{M})} |\tau|}{|f(\tau)| / \sum_{\tau \in \mathcal{F}(\mathcal{M})} |f(\tau)|}. \qquad (20)$$

A mapping $f$ is area-preserving if $\mathcal{R}_f(\tau) = 1$, for every $\tau \in \mathcal{F}(\mathcal{M})$.

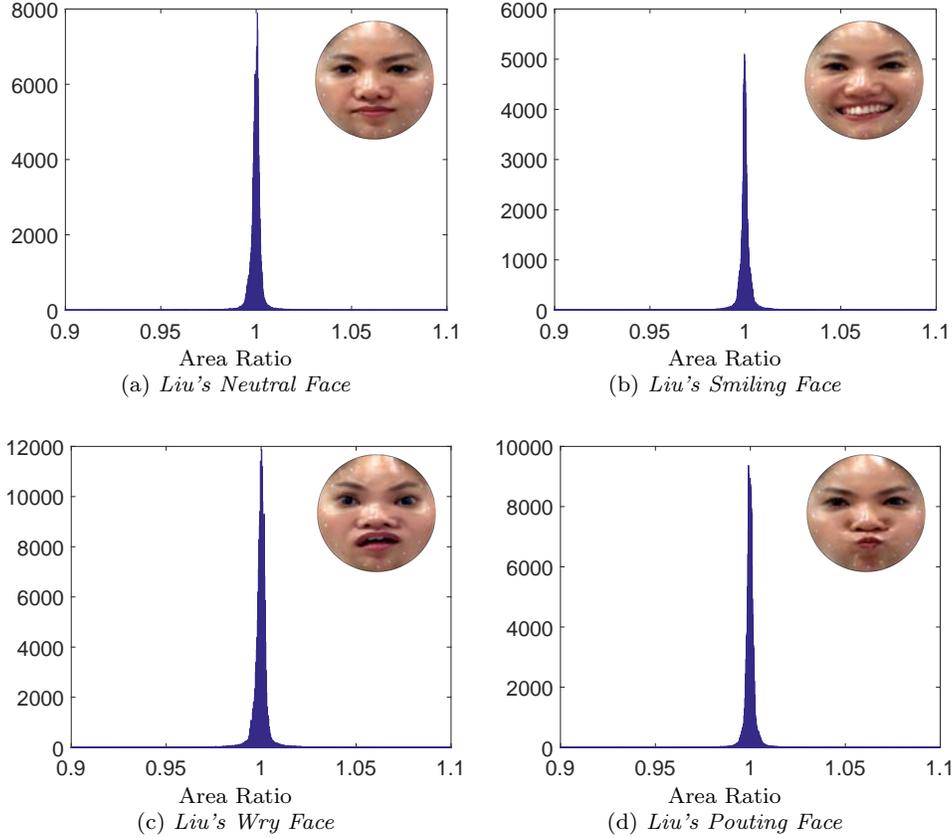

**Fig. 7** The histograms of the local area ratio of the equiareal parameterizations obtained by SEM algorithm for (a) *Liu's Neutral Face*, (b) *Liu's Smiling Face*, (c) *Liu's Wry Face* and (d) *Liu's Pouting Face*. Note that the area ratio refers as to the ratio between the area of the triangular face on the mesh model and on the disk.

Fig. 7 shows the histograms of the local area ratio of the parameterizations obtained by SEM algorithm for (a) *Liu's Neutral Face*, (b) *Liu's Smiling Face*, (c) *Liu's Wry Face* and (d) *Liu's Pouting Face*. The area ratio is defined as in (20), which measures the ratio between the area of the triangular face on the mesh model and on the disk. As shown in Fig. 7, the local area ratios of most triangular faces are close to 1, which is quite satisfactory.

Comparisons of the area distortion in terms of the total area distortion and the local area ratio for the parameterization mappings are demonstrated in Table 2 and Table 3, respectively. Also, Fig. 8 indicates that among three algorithms, the total area distortion (19) produced by SEM algorithm is always the smallest. Similarly, Figs. 9 and 10 indicate that SEM algorithm has a better accuracy than that of FSM and OMT. Note that for every demonstrated mesh model, the total area distortion produced by SEM algorithm is always less than 5%, which is satisfied in practical applications.

Figs. 9 and 10 indicate that most of the local area ratios (20) of the parameterization mapping computed by SEM algorithm is close to 1, which means the local area distortion is small.

**Table 2** The total area distortion of the equiareal parameterizations computed by FSM, OMT, and SEM algorithms. The notation "−" means the iteration of the algorithm diverges.

| Model Name | # Faces | FSM [20] | OMT [18] | SEM |
| --- | ---: | ---: | ---: | ---: |
| Nefertiti | 562 | 0.1349 | 0.1776 | 0.0295 |
| Cowboy Hat | 4,604 | 0.0724 | 0.1892 | 0.0087 |
| Chinese Vase | 5,592 | 0.4404 | − | 0.0206 |
| Bourbon Bottle | 13,088 | − | − | 0.0449 |
| Foot | 19,966 | − | 0.3653 | 0.0413 |
| Chinese Lion | 34,421 | 0.2137 | 0.1927 | 0.0143 |
| Stanford Bunny | 65,221 | − | − | 0.0342 |
| Human Brain | 96,811 | − | 0.1875 | 0.0179 |
| Left Hand | 105,860 | − | − | 0.0401 |
| Statue of Liberty | 190,162 | − | − | 0.0259 |
| Liu's Neutral Face | 193,298 | 0.0345 | 0.0739 | 0.0037 |
| Liu's Smiling Face | 205,207 | 0.0354 | 0.0731 | 0.0040 |
| Liu's Pouting Face | 207,721 | 0.0414 | 0.0718 | 0.0039 |
| Liu's Wry Face | 208,283 | 0.0626 | 0.0749 | 0.0044 |
| Isis Statue | 374,309 | − | − | 0.0185 |
| Bimba Statue | 836,740 | − | 0.2250 | 0.0084 |
| Knit Cap Man | 1,287,579 | − | 0.2950 | 0.0102 |

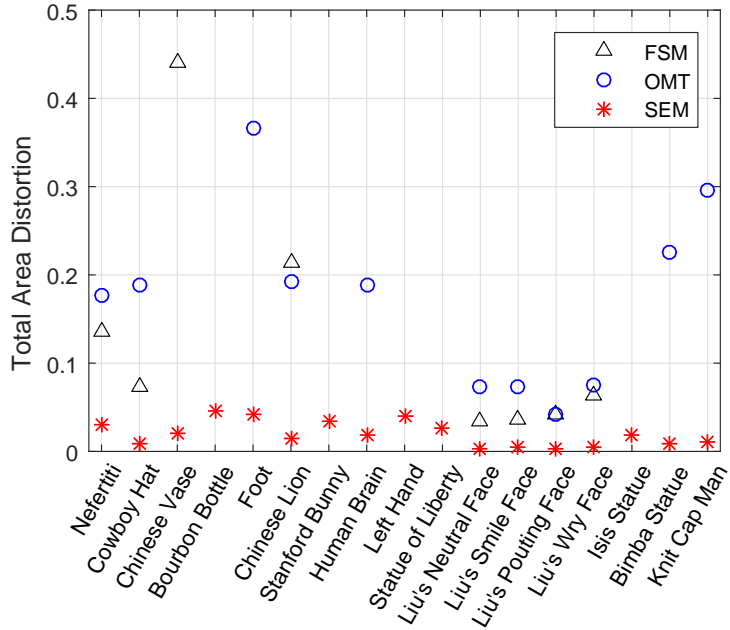

**Fig. 8** The total area distortion of the equiareal parameterizations computed by FSM, OMT, and SEM algorithms.

Note that, for every demonstrated mesh model, the iteration of SEM algorithm converges, while, for some mesh models, the iteration of FSM or OMT does not converge. It is fairly said that the proposed SEM algorithm is much more robust.

On the other hand, Figs. 11 and 12, respectively, show the relationship between the number of iterations and the area distortions as well as the stretch energy. These results indicate that the area distortion decreases while the stretch energy decreases. In other words, it is reasonable to compute an equiareal parame-terization via minimizing the stretch energy. In addition, as shown in Figs. 11 and 12, both the area distortions and the stretch energy are significantly decreased in the first three iteration steps, which shows the effectiveness of the proposed SEM algorithm.

**Table 3** The mean and SD of the local area ratio of the equiareal parameterizations by FSM [20], OMT [18] and SEM algorithms. The notation "−" means the iteration of the algorithm diverges.

| Model Name | # Faces | FSM [20] Mean | FSM [20] SD | OMT [18] Mean | OMT [18] SD | SEM Mean | SEM SD |
|---|---|---|---|---|---|---|---|
| Nefertiti | 562 | 1.0471 | 0.1319 | 1.0097 | 0.2504 | 1.0032 | 0.0598 |
| Cowboy Hat | 4,604 | 1.0449 | 0.0838 | 1.1289 | 1.5094 | 1.0000 | 0.0100 |
| Chinese Vase | 5,592 | 1.0573 | 0.8054 | − | − | 1.0496 | 0.5290 |
| Bourbon Bottle | 13,088 | − | − | − | − | 1.0116 | 0.1322 |
| Foot | 19,966 | − | − | 1.4865 | 16.3484 | 1.0092 | 0.4055 |
| Chinese Lion | 34,421 | 1.0618 | 0.2327 | 1.0712 | 1.4920 | 1.0005 | 0.0245 |
| Stanford Bunny | 65,221 | − | − | − | − | 1.0088 | 0.2718 |
| Human Brain | 96,811 | − | − | 1.3068 | 37.8400 | 1.0021 | 0.1030 |
| Left Hand | 105,860 | − | − | − | − | 1.0038 | 0.0683 |
| Statue of Liberty | 190,162 | − | − | − | − | 1.0081 | 0.5752 |
| Liu's Neutral Face | 193,298 | 1.0039 | 0.0490 | 1.0103 | 0.1058 | 1.0002 | 0.0139 |
| Liu's Smiling Face | 205,207 | 1.0045 | 0.0501 | 1.0101 | 0.1038 | 1.0002 | 0.0139 |
| Liu's Pouting Face | 207,721 | 1.0054 | 0.0555 | 1.0102 | 0.1052 | 1.0002 | 0.0143 |
| Liu's Wry Face | 208,283 | 1.0109 | 0.0747 | 1.0103 | 0.1060 | 1.0002 | 0.0148 |
| Isis Statue | 374,309 | − | − | − | − | 1.0013 | 0.1032 |
| Bimba Statue | 836,740 | − | − | 1.8758 | 267.3992 | 1.0016 | 0.4303 |
| Knit Cap Man | 1,287,579 | − | − | 1.5069 | 66.8598 | 1.0018 | 0.1397 |

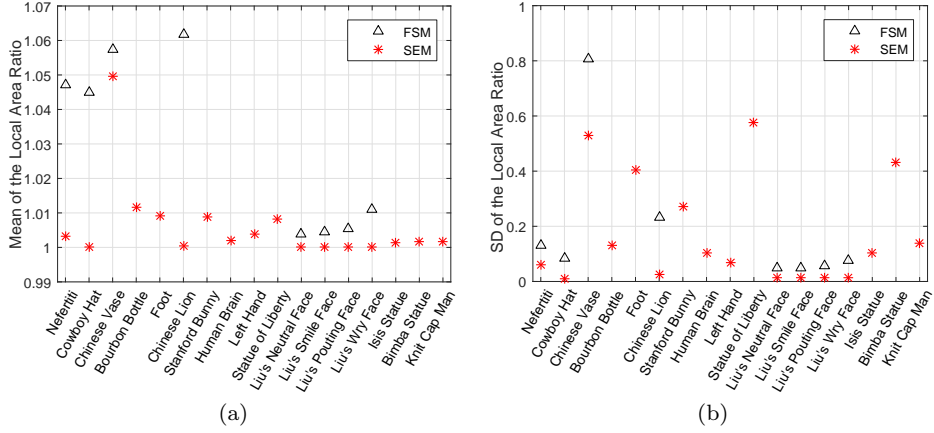

**Fig. 9** The (a) mean and (b) standard deviation of the local area ratio of the parameterizations by FSM and SEM algorithms.

## 7 Applications

In this section, we demonstrate two sample applications of the proposed SEM algorithm, namely, surface remeshing in Sect. 7.1 and surface registration in Sect. 7.2.

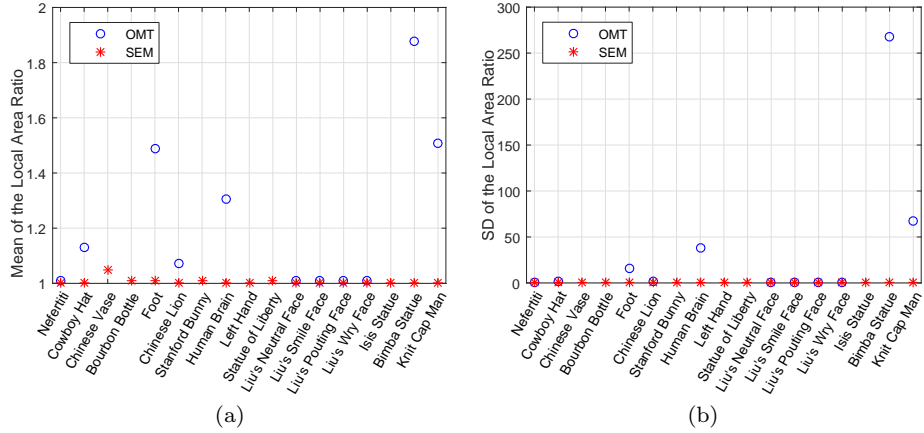

**Fig. 10** The (a) mean and (b) standard deviation of the local area ratio of the parameterizations by OMT and SEM algorithms.

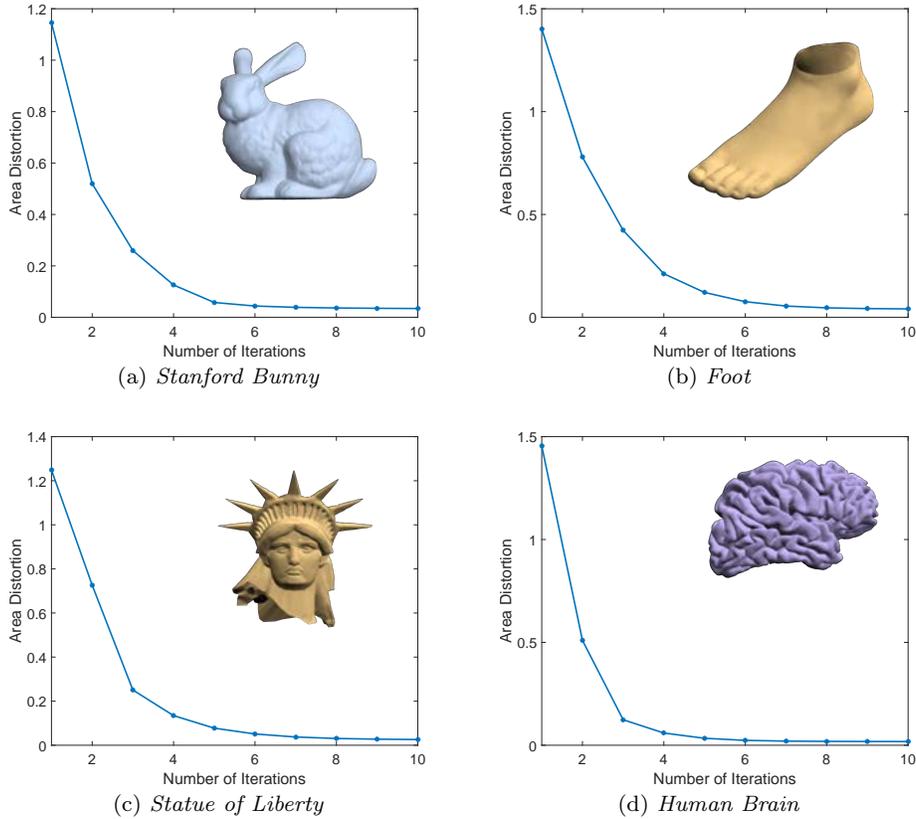

**Fig. 11** The relationship between the number of iterations and the area distortions of the parameterization obtained by SEM algorithm for (a) *Stanford Bunny*, (b) *Foot*, (c) *Statue of Liberty* and (d) *Human Brain*.

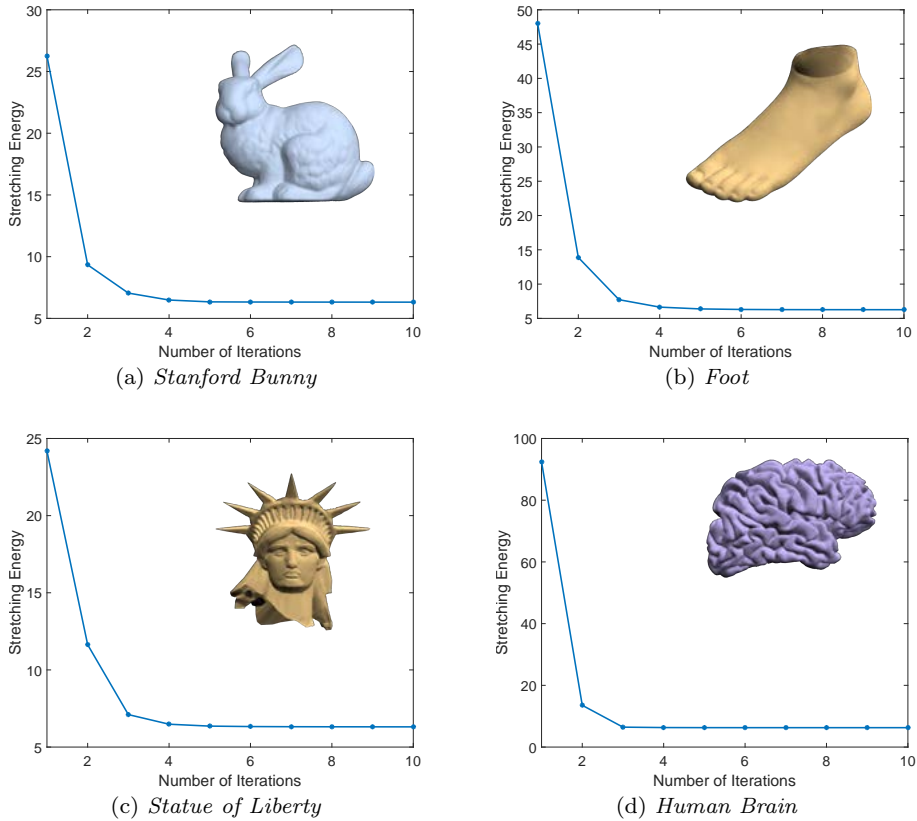

**Fig. 12** The relationship between the number of iterations and the stretch energy of the parameterization obtained by SEM algorithm for (a) *Stanford Bunny*, (b) *Foot*, (c) *Statue of Liberty* and (d) *Human Brain*.

7.1 Surface Remeshing

Surface remeshing refers as to the improvement process of mesh quality in terms of vertex sampling, regularity and triangle quality [6]. In the demonstrated numerical experiments, the raw mesh data of human faces are captured by the structured-light 3D scanner *GeoVideo* (GI company) in the ST Yau Center in Taiwan. In the raw mesh data of a human face $\mathcal{M}$, shown in Fig. 13 (a), the sizes of triangles are very different. The equiareal parameterization can be applied to the surface remeshing so that the vertex sampling on $\mathcal{M}$ becomes uniform and the mesh becomes regular after the remeshing process.

The remeshing procedure is described as the follows. First, a bijective equiareal mapping $f : \mathcal{M} \to \mathbb{D} \subset \mathbb{C}$ is computed by using the SEM algorithm. Then a regular mesh $\mathcal{U}$ of a unit disk with a uniform sampling is covered on the image $f(\mathcal{M})$. Finally, the remeshed surface $f^{-1}(\mathcal{U})$ is obtained by the one-to-one correspondences between the barycentric coordinates of $\mathcal{M}$ and $f(\mathcal{M})$.

Fig. 13 shows a comparison of the raw mesh data and the remeshed data of a human face before and after the remeshing process, respectively. Some obtuse triangles at the nose part disappear after the process. Fig. 14 shows the histograms of the areas of triangles of the raw mesh data and the remeshed data of the human face, which indicate that the sampling of vertices becomes more uniform after the remeshing process.

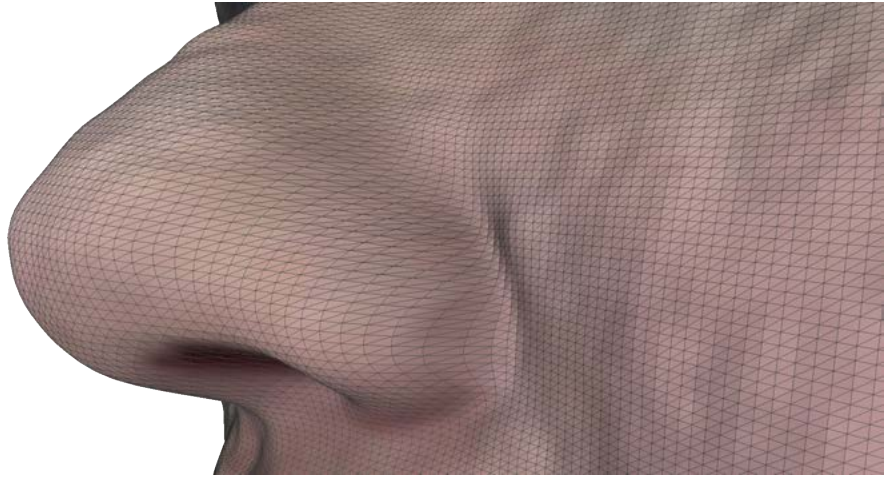

(a) The raw mesh data.

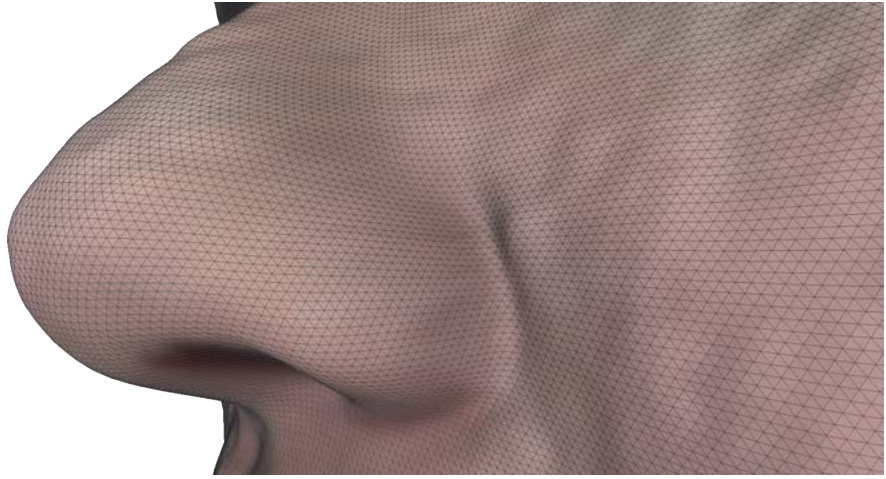

(b) The remeshed data.

**Fig. 13** A comparison of (a) the raw mesh data and (b) the remeshed data of the human face.

7.2 Surface Registration

Surface registration has been widely applied to geometry processing tasks [19, 21]. Given two meshes $\mathcal{M}$ and $\mathcal{N}$ with $m$ and $n$ vertices, respectively, and a set of landmark pairs
$$\{(\mathbf{p}_\ell, \mathbf{q}_\ell) \,|\, \mathbf{p}_\ell \in \mathcal{M}, \mathbf{q}_\ell \in \mathcal{N}\}_{\ell=1}^d.$$
The goal of the surface registration [19] is to construct a smooth bijective mapping $h : \mathcal{M} \to \mathcal{N}$ such that $h(\mathbf{p}_\ell) = \mathbf{q}_\ell$. With the help of the parameterization mapping $g : \mathcal{N} \to \mathbb{D}$, the surface registration problem in the three-dimensional space is reduced into a planar registration problem on a unit disk. The reduced problem is to find a smooth bijective mapping $h : \mathcal{M} \to g(\mathcal{N}) \subset \mathbb{D}$ such that $h(\mathbf{p}_\ell) = g(\mathbf{q}_\ell)$. In this section, we propose an efficient algorithm for computing the registration mappings between surfaces.

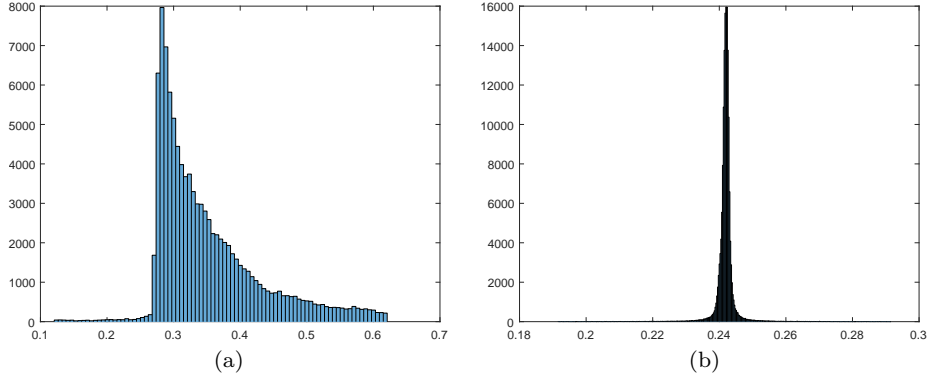

**Fig. 14** The histograms of the areas of triangles of (a) the raw mesh data and (b) the remeshed data of the human face.

Suppose the discrete conformal mappings $f : \mathcal{M} \to \mathbb{D}$ and $g : \mathcal{N} \to \mathbb{D}$ are computed by using the CEM algorithm [22]. We denote the discrete mappings by the vectors $\mathbf{f} \in \mathbb{C}^m$ and $\mathbf{g} \in \mathbb{C}^n$, respectively. Let P and Q be the ordered sets of indices of the landmarks on $\mathcal{M}$ and $\mathcal{N}$, respectively, i.e.,

$$f(\mathbf{p}_\ell) = (\mathbf{f}_\mathtt{P})_\ell$$

and

$$g(\mathbf{q}_\ell) = (\mathbf{g}_\mathtt{Q})_\ell.$$

A smooth registration mapping (SRM) $h : \mathcal{M} \to g(\mathcal{N})$ can be obtained by minimizing the *registration energy* defined as

$$\mathcal{E}_R(\mathbf{h}) = \|L(\mathbf{h})\,\mathbf{h}\|_2^2 + \sum_{\ell=1}^d \lambda_\ell^2 |(\mathbf{h}_\mathtt{P})_\ell - (\mathbf{g}_\mathtt{Q})_\ell|^2 \qquad (21)$$

in which $\lambda_\ell$ is an appropriate weight for the landmark pair $((\mathbf{h}_\mathtt{P})_\ell, (\mathbf{g}_\mathtt{Q})_\ell)$, for $\ell = 1, \ldots, d$, $L(\mathbf{h})$ is the Laplacian matrix similar as in (6) defined by

$$[L(\mathbf{h})]_{i,j} = \begin{cases} -\omega_{i,j}(\mathbf{h}) & \text{if } [\mathbf{v}_i, \mathbf{v}_j] \in \mathcal{E}(\mathcal{M}), \\ \sum_{\ell \neq i} \omega_{i,\ell}(\mathbf{h}) & \text{if } j = i, \\ 0 & \text{otherwise,} \end{cases} \qquad (22)$$

where $\omega_{i,j}(\mathbf{h})$ is the cotangent weight defined as

$$\omega_{i,j}(\mathbf{h}) = \frac{\cot(\alpha_{i,j}(\mathbf{h})) + \cot(\alpha_{j,i}(\mathbf{h}))}{2}$$

with $\alpha_{i,j}(\mathbf{h})$ and $\alpha_{j,i}(\mathbf{h})$ being two angles opposite to the edge $[\mathbf{h}_i, \mathbf{h}_j]$ connecting vertices $\mathbf{h}_i$ and $\mathbf{h}_j$ on $\mathbb{C}$.

The surface registration process can be performed as follows. First, an initial mapping $\mathbf{h}^{(0)}$ can be computed by the composition of the conformal mapping $f$ and a Möbius transformation such that $(\mathbf{h}_\mathtt{P}^{(0)})_\ell = (\mathbf{g}_\mathtt{Q})_\ell$ for some $\ell$, i.e.,

$$(\mathbf{h}^{(0)})_j = \frac{\mathbf{f}_j - ((\mathbf{f}_\mathtt{P})_\ell - (\mathbf{g}_\mathtt{Q})_\ell)}{1 - \mathbf{f}_j((\mathbf{f}_\mathtt{P})_\ell - (\mathbf{g}_\mathtt{Q})_\ell)^*},$$

for $j = 1, \ldots, m$. Then the registration energy (21) can be minimized by the iterative procedure

$$\mathbf{h}^{(k+1)} = \operatorname*{argmin}_{\mathbf{h}_\mathtt{B}^{(k+1)} = \mathbf{h}_\mathtt{B}^{(0)}} \left( \left\| L(\mathbf{h}^{(k)}) \, \mathbf{h} \right\|_2^2 + \| \Lambda(\mathbf{h}_\mathtt{P} - \mathbf{g}_\mathtt{Q}) \|_2^2 \right), \tag{23}$$

where $\Lambda$ is the diagonal matrix with the diagonal entry $\Lambda_{\ell,\ell} = \lambda_\ell$, for $\ell = 1, \ldots, d$. Let $P$ be the matrix defined by

$$P_{i,j} = \begin{cases} 1 & \text{if } \mathtt{P}_i = j, \\ 0 & \text{otherwise,} \end{cases}$$

where $\mathtt{P}_i$ is the $i$-th element of the ordered set $\mathtt{P}$. Then Eq. (23) can be solved by the least squared method

$$A^{(k)} \, \mathbf{h}_\mathtt{I}^{(k+1)} = \mathbf{b}^{(k)} \tag{24}$$

in which

$$A^{(k)} = \left[ L(\mathbf{h}^{(k)}) \right]_{\mathtt{I},\mathtt{I}}^\top \left[ L(\mathbf{h}^{(k)}) \right]_{\mathtt{I},\mathtt{I}} + P_{\cdot,\mathtt{I}}^\top P_{\cdot,\mathtt{I}}$$

and

$$\mathbf{b}^{(k)} = - \left[ L(\mathbf{h}^{(k)}) \right]_{\mathtt{I},\mathtt{I}}^\top \left[ L(\mathbf{h}^{(k)}) \right]_{\mathtt{I},\mathtt{B}} \mathbf{h}_\mathtt{B}^{(0)} + P_{\cdot,\mathtt{I}}^\top \Lambda \mathbf{g}_\mathtt{Q}.$$

The algorithm for computing the SRM is summarized in Algorithm 2.

Fig. 15 (a), (c) and (e) show the models of *Lin's Neutral Face*, *Lin's Sad Face* and *Lin's Pouting Face*, respectively. Fig. 15 (b) shows the conformal parameterization of the model of *Lin's Neutral Face* computed using the CEM algorithm [22]. Fig. 15 (d) and (f) show the registration mappings of *Lin's Sad Face* and *Lin's Pouting Face* computed by SRM algorithm, respectively, where the target surface is *Lin's Neutral Face*. The checkerboard patterns in Fig. 15 show the correspondence between each surface via the computed registration mappings.

It is remarkable that the demonstrated registration mappings computed by SRM algorithm are bijective, and the distortion of the mapping between the selected landmark pairs is close to zero.

A demo video of surface morphing between facial expressions via the registration mappings can be found at https://youtu.be/BjAL-pN44k4.

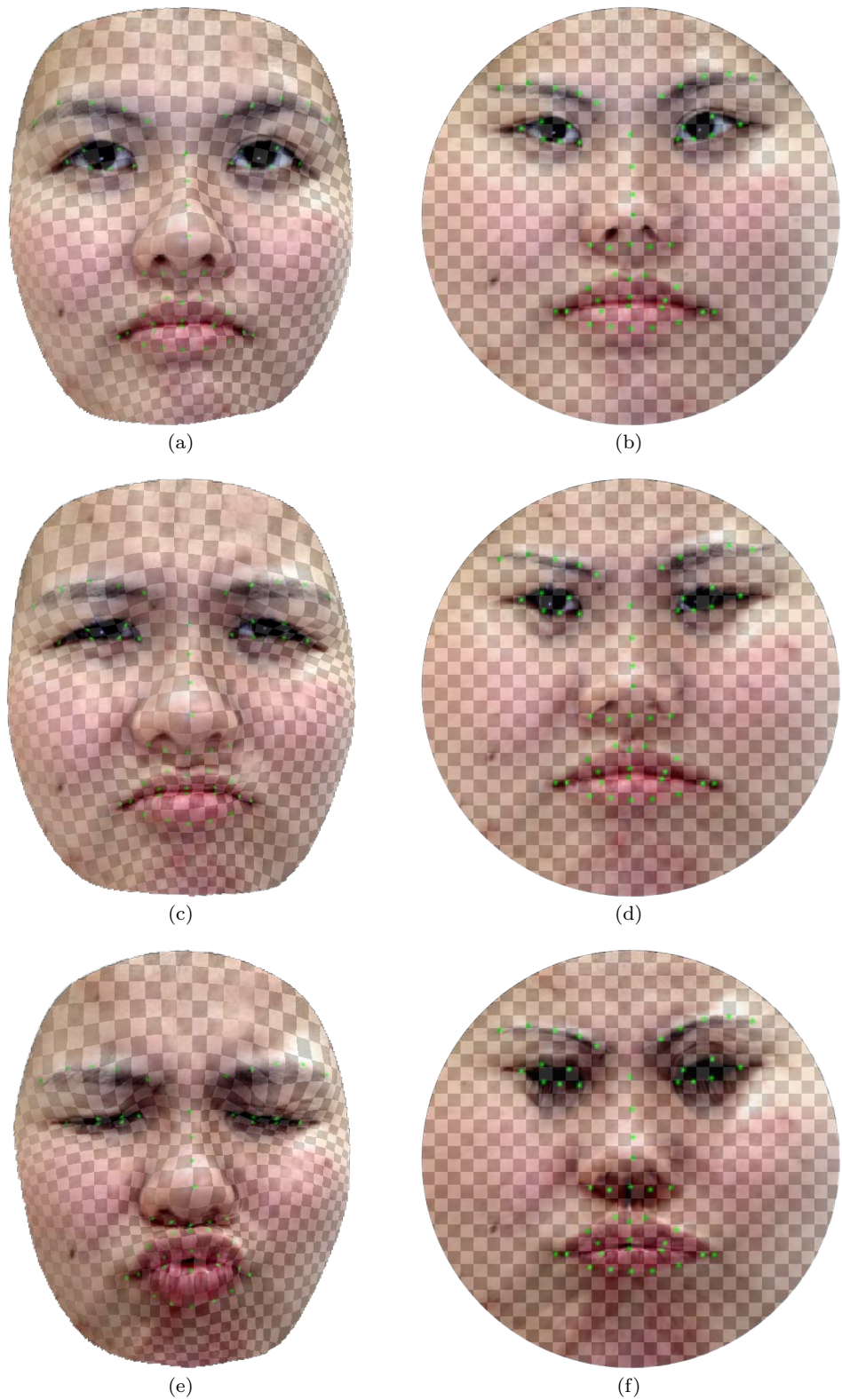

**Fig. 15** (a) The model of *Lin's Neutral Face*. (b) The conformal parameterization obtained by CEM algorithm. (c) The model of *Lin's Sad Face*. (d) The registration mapping obtained by SRM algorithm. (e) The model of *Lin's Pouting Face*. (f) The registration mapping obtained by SRM algorithm.

**Algorithm 2** Smooth Registration Mapping (SRM)

---

**Input:** Two simply connected open meshes $\mathcal{M}$ and $\mathcal{N}$, two index sets of landmark pairs P and Q, and the weights for the landmark pairs $\lambda_\ell$, for $\ell = 1, \ldots, d$.

**Output:** A pair of registration mappings $g : \mathcal{N} \to \mathbb{D}$ and $h : \mathcal{M} \to \mathbb{D}$, represented by **g** and **h**, respectively, such that $\mathbf{h}_\mathtt{P} \approx \mathbf{g}_\mathtt{Q}$.

1: Compute the conformal parameterizations **f** and **g** of the meshes $\mathcal{M}$ and $\mathcal{N}$, respectively, using the CEM Algorithm [22].

2: Möbius transformation: $\mathbf{h}_j \leftarrow \frac{\mathbf{f}_j - ((\mathbf{f}_\mathtt{P})_\ell - (\mathbf{g}_\mathtt{Q})_\ell)}{1 - \mathbf{f}_j((\mathbf{f}_\mathtt{P})_\ell - (\mathbf{g}_\mathtt{Q})_\ell)^*}$, for some $\ell$.

3: Construct the matrix $P$ with
$$P_{i,j} = \begin{cases} 1 & \text{if } \mathtt{P}_i = j, \\ 0 & \text{otherwise.} \end{cases}$$

4: **while** not convergent **do**

5:    Construct the matrix
$$A = [L(\mathbf{h})]_{\mathtt{I},\mathtt{I}}^\top [L(\mathbf{h})]_{\mathtt{I},\mathtt{I}} + P_{\cdot,\mathtt{I}}^\top P_{\cdot,\mathtt{I}}$$

6:    Construct the vector
$$\mathbf{b} = -[L(\mathbf{h})]_{\mathtt{I},\mathtt{I}}^\top [L(\mathbf{h})]_{\mathtt{I},\mathtt{B}} \mathbf{h}_\mathtt{B} + P_{\cdot,\mathtt{I}}^\top \mathrm{diag}\left((\lambda_1, \ldots, \lambda_{n_\mathtt{P}})^\top\right) \mathbf{g}_\mathtt{Q}.$$

7:    Update $\mathbf{h}_\mathtt{I}$ by solving the linear system
$$A\, \mathbf{h}_\mathtt{I} = \mathbf{b}.$$

8: **end while**

---

## 8 Conclusion

In this paper, we propose a novel efficient SEM algorithm for the computation of equiareal parameterizations of simply connected open surfaces. The demonstrated numerical results of the SEM algorithm has the following advantages:

- **Highly Improved Robustness**: For every demonstrated mesh models, the iteration of the SEM algorithm converges and produces a reasonably good equiareal parameterization mapping.
- **Highly Improved Efficiency**: The efficiency of the proposed SEM algorithm outperforms other state-of-the-art algorithms [20,18]. Numerical results indicate that an equiareal parameterization can be computed in less than 3 seconds for a mesh model of more than 100,000 triangular faces by SEM algorithm.
- **Small Area Distortion**: For every demonstrated numerical result, the total area distortion is less than 5%, which outperforms other state-of-the-art algorithms [20,18].

Thanks to the reliablity and the small area distortion of the SEM algorithm, the computation of the surface remeshing and surface registration can be performed efficiently and robustly.

**Acknowledgements**  The authors want to thank Prof. Xianfeng David Gu for the useful discussion and the executable program files of the OMT algorithm. They also thank the ST Yau Center at NCTU, Taiwan for the support on 3D scanner and camera devices. This work is partially supported by the Ministry of Science and Technology, the National Center for Theoretical Sciences, the Taida

Institute for Mathematical Sciences, and the ST Yau Center in Taiwan. This work is supported in part by the Ministry of Science and Technology, Taiwan, Republic of China, under Grant MOST 106-2811-M-009-046.

## A Consistency of the Stretch Factor

**Theorem 3** *Given a mesh $\mathcal{M}$ and a parameterization mapping $f : \mathcal{M} \to \mathcal{D} \subset \mathbb{C}$. The local stretch factor, defined as (5), with respect to $\mathbf{f} = (f(\mathbf{v}_1), \ldots, f(\mathbf{v}_n))^\top$ satisfies*

$$\sigma_{\mathbf{f}}([\mathbf{v}_i, \mathbf{v}_j, \mathbf{v}_k])^2 = \det\left(I_{f^{-1}}|_{[\mathbf{f}_i, \mathbf{f}_j, \mathbf{f}_k]}\right), \tag{25}$$

*for every $[\mathbf{v}_i, \mathbf{v}_j, \mathbf{v}_k] \in \mathcal{F}(\mathcal{M})$.*

*Proof* Note that the partial derivatives are translation invariant. Without loss of generality, let $\mathbf{v}_i + \mathbf{v}_j + \mathbf{v}_k = 0$, and $\mathbf{f}_i + \mathbf{f}_j + \mathbf{f}_k = 0$. That is, $\mathbf{v}_k = -\mathbf{v}_i - \mathbf{v}_j$, and $\mathbf{f}_k = -\mathbf{f}_i - \mathbf{f}_j$. A direct computation yields that

$$|[\mathbf{f}_i, \mathbf{f}_j, \mathbf{f}_k]| = \frac{3}{2}|\mathbf{f}_i^1 \mathbf{f}_j^2 - \mathbf{f}_i^2 \mathbf{f}_j^1|.$$

Similarly,

$$\begin{aligned}
|[\mathbf{v}_i, \mathbf{v}_j, \mathbf{v}_k]|^2 &= \frac{1}{4}\|(\mathbf{v}_i - \mathbf{v}_k) \times (\mathbf{v}_j - \mathbf{v}_k)\|^2 \\
&= \frac{1}{4}\|(2\mathbf{v}_i + \mathbf{v}_j) \times (\mathbf{v}_i + 2\mathbf{v}_j)\|^2 \\
&= \frac{1}{4}\left(\|2\mathbf{v}_i + \mathbf{v}_j\|^2\|\mathbf{v}_i + 2\mathbf{v}_j\|^2 - ((2\mathbf{v}_i + \mathbf{v}_j)^\top(\mathbf{v}_i + 2\mathbf{v}_j))^2\right) \\
&= \frac{9}{4}\left((\mathbf{v}_i^\top \mathbf{v}_i)(\mathbf{v}_j^\top \mathbf{v}_j) - (\mathbf{v}_i^\top \mathbf{v}_j)^2\right).
\end{aligned}$$

On the other hand,

$$\frac{\partial f^{-1}|_{[\mathbf{f}_i, \mathbf{f}_j, \mathbf{f}_k]}}{\partial z^1} = \frac{\mathbf{f}_j^2 \mathbf{v}_i - \mathbf{f}_i^2 \mathbf{v}_j}{|\mathbf{f}_i^1 \mathbf{f}_j^2 - \mathbf{f}_i^2 \mathbf{f}_j^1|},$$

and

$$\frac{\partial f^{-1}|_{[\mathbf{f}_i, \mathbf{f}_j, \mathbf{f}_k]}}{\partial z^2} = \frac{\mathbf{f}_i^1 \mathbf{v}_j - \mathbf{f}_j^1 \mathbf{v}_i}{|\mathbf{f}_i^1 \mathbf{f}_j^2 - \mathbf{f}_i^2 \mathbf{f}_j^1|}.$$

Hence,

$$\begin{aligned}
a_{\mathbf{f}}|_{[\mathbf{f}_i, \mathbf{f}_j, \mathbf{f}_k]} &= \left(\frac{\partial f^{-1}|_{[\mathbf{f}_i, \mathbf{f}_j, \mathbf{f}_k]}}{\partial z^1}\right)^\top \left(\frac{\partial f^{-1}|_{[\mathbf{f}_i, \mathbf{f}_j, \mathbf{f}_k]}}{\partial z^1}\right) \\
&= \frac{(\mathbf{f}_j^2)^2(\mathbf{v}_i^\top \mathbf{v}_i) - 2\mathbf{f}_1^2 \mathbf{f}_j^2(\mathbf{v}_i^\top \mathbf{v}_j) + (\mathbf{f}_i^2)^2(\mathbf{v}_j^\top \mathbf{v}_j)}{|\mathbf{f}_i^1 \mathbf{f}_j^2 - \mathbf{f}_i^2 \mathbf{f}_j^1|},
\end{aligned}$$

$$\begin{aligned}
b_{\mathbf{f}}|_{[\mathbf{f}_i, \mathbf{f}_j, \mathbf{f}_k]} &= \left(\frac{\partial f^{-1}|_{[\mathbf{f}_i, \mathbf{f}_j, \mathbf{f}_k]}}{\partial z^1}\right)^\top \left(\frac{\partial f^{-1}|_{[\mathbf{f}_i, \mathbf{f}_j, \mathbf{f}_k]}}{\partial z^2}\right) \\
&= \frac{-\mathbf{f}_j^1 \mathbf{f}_j^2(\mathbf{v}_i^\top \mathbf{v}_i) + (\mathbf{f}_i^1 \mathbf{f}_j^2 + \mathbf{f}_i^2 \mathbf{f}_j^1)(\mathbf{v}_i^\top \mathbf{v}_j) - \mathbf{f}_i^1 \mathbf{f}_i^2(\mathbf{v}_j^\top \mathbf{v}_j)}{|\mathbf{f}_i^1 \mathbf{f}_j^2 - \mathbf{f}_i^2 \mathbf{f}_j^1|},
\end{aligned}$$

and

$$\begin{aligned}
c_{\mathbf{f}}|_{[\mathbf{f}_i, \mathbf{f}_j, \mathbf{f}_k]} &= \left(\frac{\partial f^{-1}|_{[\mathbf{f}_i, \mathbf{f}_j, \mathbf{f}_k]}}{\partial z^2}\right)^\top \left(\frac{\partial f^{-1}|_{[\mathbf{f}_i, \mathbf{f}_j, \mathbf{f}_k]}}{\partial z^2}\right) \\
&= \frac{(\mathbf{f}_j^1)^2(\mathbf{v}_i^\top \mathbf{v}_i) - 2\mathbf{f}_1^1 \mathbf{f}_j^1(\mathbf{v}_i^\top \mathbf{v}_j) + (\mathbf{f}_i^1)^2(\mathbf{v}_j^\top \mathbf{v}_j)}{|\mathbf{f}_i^1 \mathbf{f}_j^2 - \mathbf{f}_i^2 \mathbf{f}_j^1|}.
\end{aligned}$$

It follows that

$$\det\left(I_{f^{-1}}|_{[\mathbf{f}_i,\mathbf{f}_j,\mathbf{f}_k]}\right) = a_{\mathbf{f}}(z)c_{\mathbf{f}}(z) - b_{\mathbf{f}}(z)^2$$

$$= \frac{(\mathbf{f}_i^1\mathbf{f}_j^2 - \mathbf{f}_i^2\mathbf{f}_j^1)^2\left((\mathbf{v}_i^\top\mathbf{v}_i)(\mathbf{v}_j^\top\mathbf{v}_j) - (\mathbf{v}_i^\top\mathbf{v}_j)^2\right)}{|\mathbf{f}_i^1\mathbf{f}_j^2 - \mathbf{f}_i^2\mathbf{f}_j^1|^4}$$

$$= \frac{(\mathbf{v}_i^\top\mathbf{v}_i)(\mathbf{v}_j^\top\mathbf{v}_j) - (\mathbf{v}_i^\top\mathbf{v}_j)^2}{|\mathbf{f}_i^1\mathbf{f}_j^2 - \mathbf{f}_i^2\mathbf{f}_j^1|^2}$$

$$= \frac{\frac{9}{4}\left((\mathbf{v}_i^\top\mathbf{v}_i)(\mathbf{v}_j^\top\mathbf{v}_j) - (\mathbf{v}_i^\top\mathbf{v}_j)^2\right)}{(\frac{3}{2}|\mathbf{f}_i^1\mathbf{f}_j^2 - \mathbf{f}_i^2\mathbf{f}_j^1|)^2}$$

$$= \frac{|[\mathbf{v}_i,\mathbf{v}_j,\mathbf{v}_k]|^2}{|[\mathbf{f}_i,\mathbf{f}_j,\mathbf{f}_k]|^2} = \sigma_{\mathbf{f}}([\mathbf{v}_i,\mathbf{v}_j,\mathbf{v}_k])^2.$$

□

## B Derivations of Gradients of the Stretch Energy

In this section, we derive the explicit formulas (9) and (10) for the gradients of the stretch energy.

**Proposition 1** *Given a mesh $\mathcal{M}$ of $n$ vertices and a bijective piecewise affine mapping $f: \mathcal{M} \to \mathcal{D} \subset \mathbb{C}$. Let $\mathbf{f} = (f(\mathbf{v_1}),\ldots,f(\mathbf{v_n}))^\top$ and $L(\mathbf{f})$ be defined as (6). Suppose $[\mathbf{v}_j,\mathbf{v}_k] \in \mathcal{E}(\mathcal{M})$. Then Eq. (9) holds, for $s = 1,2$.*

*Proof* For each edge $[\mathbf{v}_j,\mathbf{v}_k] \in \mathcal{E}(\mathcal{M})$,

$$\left[(\mathbf{f}^s)^\top \frac{\partial L(\mathbf{f})}{\partial \mathbf{f}^s}\right]_{j,k} = \sum_{\alpha=1}^n \mathbf{f}_\alpha^s \frac{\partial L_{\alpha,j}}{\partial \mathbf{f}_k^s} = \sum_{\alpha \neq j,k} \mathbf{f}_\alpha^s \frac{\partial L_{\alpha,j}}{\partial \mathbf{f}_k^s} + \mathbf{f}_j^s \frac{\partial L_{j,j}}{\partial \mathbf{f}_k^s} + \mathbf{f}_k^s \frac{\partial L_{k,j}}{\partial \mathbf{f}_k^s}$$

$$= \sum_{\alpha \neq j,k} (\mathbf{f}_\alpha^s - \mathbf{f}_j^s)\frac{\partial L_{\alpha,j}}{\partial \mathbf{f}_k^s} + (\mathbf{f}_k^s - \mathbf{f}_j^s)\frac{\partial L_{k,j}}{\partial \mathbf{f}_k^s}$$

$$= -\frac{1}{2}\sum_{[\mathbf{v}_\alpha,\mathbf{v}_j,\mathbf{v}_k]\in\mathcal{F}(\mathcal{M})}(\mathbf{f}_\alpha^s - \mathbf{f}_j^s)\frac{2\mathbf{f}_k^s - \mathbf{f}_\alpha^s - \mathbf{f}_j^s}{2|[\mathbf{v}_i,\mathbf{v}_j,\mathbf{v}_k]|}$$

$$\quad - \frac{1}{2}(\mathbf{f}_k^s - \mathbf{f}_j^s)\sum_{[\mathbf{v}_\alpha,\mathbf{v}_j,\mathbf{v}_k]\in\mathcal{F}(\mathcal{M})}\frac{\mathbf{f}_j^s - \mathbf{f}_\alpha^s}{2|[\mathbf{v}_\alpha,\mathbf{v}_j,\mathbf{v}_k]|}$$

$$= -\frac{1}{2}\sum_{[\mathbf{v}_\alpha,\mathbf{v}_j,\mathbf{v}_k]\in\mathcal{F}(\mathcal{M})}\frac{(\mathbf{f}_k^s - \mathbf{f}_\alpha^s)(\mathbf{f}_\alpha^s - \mathbf{f}_j^s)}{2|[\mathbf{v}_\alpha,\mathbf{v}_j,\mathbf{v}_k]|}.$$

On the other hand, the diagonal entries

$$\left[(\mathbf{f}^s)^\top \frac{\partial L(\mathbf{f})}{\partial \mathbf{u}^s}\right]_{j,j} = \sum_{\alpha=1}^n \mathbf{f}_\alpha^s \frac{\partial L_{\alpha,j}}{\partial \mathbf{f}_j^s} = \sum_{\alpha \neq j} \mathbf{f}_\alpha^s \frac{\partial L_{\alpha,j}}{\partial \mathbf{f}_j^s} + \mathbf{f}_j^s \frac{\partial L_{j,j}}{\partial \mathbf{f}_j^s} = \sum_{\alpha \neq j}(\mathbf{f}_\alpha^s - \mathbf{f}_j^s)\frac{\partial L_{\alpha,j}}{\partial \mathbf{f}_j^s}$$

$$= -\frac{1}{2}\sum_{\alpha \neq j}(\mathbf{f}_\alpha^s - \mathbf{f}_j^s)\sum_{[\mathbf{v}_\alpha,\mathbf{v}_j,\mathbf{v}_\beta]\in\mathcal{F}(\mathcal{M})}\frac{\mathbf{f}_\alpha^s - \mathbf{f}_\beta^s}{2|[\mathbf{v}_\alpha,\mathbf{v}_j,\mathbf{v}_\beta]|}$$

$$= -\frac{1}{2}\sum_{[\mathbf{v}_\alpha,\mathbf{v}_j,\mathbf{v}_\beta]\in\mathcal{F}(\mathcal{M})}\frac{(\mathbf{f}_\alpha^s - \mathbf{f}_\beta^s)^2}{2|[\mathbf{v}_\alpha,\mathbf{v}_j,\mathbf{v}_\beta]|}.$$

□

**Proposition 2** *Given a mesh $\mathcal{M}$ of $n$ vertices and a bijective piecewise affine mapping $f : \mathcal{M} \to \mathcal{D} \subset \mathbb{C}$. Let $\mathbf{f} = (f(\mathbf{v_1}), \ldots, f(\mathbf{v_n}))^\top$ and $L(\mathbf{f})$ be defined as (6). Suppose $[\mathbf{v}_j, \mathbf{v}_k] \in \mathcal{E}(\mathcal{M})$. Then Eq. (10) holds, for $(s,t) = (1,2), (2,1)$.*

*Proof* For each edge $[\mathbf{v}_j, \mathbf{v}_k] \in \mathcal{E}(\mathcal{M})$,

$$\left[(\mathbf{f}^s)^\top \frac{\partial L(\mathbf{f})}{\partial \mathbf{f}^t}\right]_{j,k} = \sum_{\alpha=1}^n \mathbf{f}_\alpha^s \frac{\partial L_{\alpha,j}}{\partial \mathbf{f}_k^t} = \sum_{\alpha \neq j,k} \mathbf{f}_\alpha^s \frac{\partial L_{\alpha,j}}{\partial \mathbf{f}_k^t} + \mathbf{f}_j^s \frac{\partial L_{j,j}}{\partial \mathbf{f}_k^t} + \mathbf{f}_k^s \frac{\partial L_{k,j}}{\partial \mathbf{f}_k^t}$$

$$= \sum_{\alpha \neq j,k} (\mathbf{f}_\alpha^s - \mathbf{f}_j^s) \frac{\partial L_{\alpha,j}}{\partial \mathbf{f}_k^s} + (\mathbf{f}_k^s - \mathbf{f}_j^s) \frac{\partial L_{k,j}}{\partial \mathbf{f}_k^s}$$

$$= -\frac{1}{2} \sum_{[\mathbf{v}_\alpha, \mathbf{v}_j, \mathbf{v}_k] \in \mathcal{F}(\mathcal{M})} (\mathbf{f}_\alpha^s - \mathbf{f}_j^s) \frac{2\mathbf{f}_k^t - \mathbf{f}_\alpha^t - \mathbf{f}_j^t}{2|[\mathbf{v}_i, \mathbf{v}_j, \mathbf{v}_k]|}$$

$$- \frac{1}{2}(\mathbf{f}_k^s - \mathbf{f}_j^s) \sum_{[\mathbf{v}_\alpha, \mathbf{v}_j, \mathbf{v}_k] \in \mathcal{F}(\mathcal{M})} \frac{\mathbf{f}_j^t - \mathbf{f}_\alpha^t}{2|[\mathbf{v}_\alpha, \mathbf{v}_j, \mathbf{v}_k]|}$$

$$= -\frac{1}{2} \sum_{[\mathbf{v}_\alpha, \mathbf{v}_j, \mathbf{v}_k] \in \mathcal{F}(\mathcal{M})} \frac{2(\mathbf{f}_k^s - \mathbf{f}_\alpha^s)(\mathbf{f}_\alpha^t - \mathbf{f}_j^t) - (\mathbf{f}_j^s - \mathbf{f}_\alpha^s)(\mathbf{f}_\alpha^t - \mathbf{f}_k^t)}{2|[\mathbf{v}_\alpha, \mathbf{v}_j, \mathbf{v}_k]|}.$$

On the other hand, the diagonal entries

$$\left[(\mathbf{f}^s)^\top \frac{\partial L(\mathbf{f})}{\partial \mathbf{f}^t}\right]_{j,j} = \sum_{\alpha=1}^n \mathbf{f}_\alpha^s \frac{\partial L_{\alpha,j}}{\partial \mathbf{f}_j^t} = \sum_{\alpha \neq j} \mathbf{f}_\alpha^s \frac{\partial L_{\alpha,j}}{\partial \mathbf{f}_j^t} + \mathbf{f}_j^s \frac{\partial L_{j,j}}{\partial \mathbf{f}_j^t} = \sum_{\alpha \neq j} (\mathbf{f}_\alpha^s - \mathbf{f}_j^s) \frac{\partial L_{\alpha,j}}{\partial \mathbf{f}_j^t}$$

$$= -\frac{1}{2} \sum_{\alpha \neq j} (\mathbf{f}_\alpha^s - \mathbf{f}_j^s) \sum_{[\mathbf{v}_\alpha, \mathbf{v}_j, \mathbf{v}_\beta] \in \mathcal{F}(\mathcal{M})} \frac{\mathbf{f}_\alpha^t - \mathbf{f}_\beta^t}{2|[\mathbf{v}_\alpha, \mathbf{v}_j, \mathbf{v}_\beta]|}$$

$$= -\frac{1}{2} \sum_{[\mathbf{v}_\alpha, \mathbf{v}_j, \mathbf{v}_\beta] \in \mathcal{F}(\mathcal{M})} \frac{(\mathbf{f}_\alpha^s - \mathbf{f}_\beta^s)(\mathbf{f}_\alpha^t - \mathbf{f}_\beta^t)}{2|[\mathbf{v}_\alpha, \mathbf{v}_j, \mathbf{v}_\beta]|}.$$

□